\documentclass[amssymb, preprintnumbers, showpacs, showkeys, aps, pra, superscriptaddress, twocolumn, longbibliography]{revtex4-2}

\usepackage{color} 
\usepackage{slashed}
\usepackage{graphicx}
\usepackage[dvipsnames]{xcolor}
\usepackage{amsmath}
\usepackage{latexsym}
\usepackage{epstopdf}
\usepackage{amsmath}
\usepackage{enumitem}
\usepackage{amssymb,amsmath}
\usepackage{multirow}
\usepackage{mathtools}
\usepackage{bbm}
\usepackage{bm}
\usepackage{amsfonts}
\usepackage{amssymb}
\usepackage{calligra}
\usepackage{calrsfs}
\usepackage{makecell}
\usepackage[normalem]{ulem}
\usepackage[american]{babel}
\usepackage{verbatim}
\usepackage{overpic}
\usepackage{hyperref}

\begin{document}

\title{Bilayer crystals in a polar-molecules system}

\author{Vinicius Zampronio}%
\email{vpedroso@if.usp.br}
\affiliation{Dipartimento di Fisica e Astronomia, Universit\`a di Firenze, I-50019, Sesto Fiorentino (FI), Italy}
\affiliation{Instituto de Física, Universidade de São Paulo, 05508-090, São Paulo (SP), Brazil}

\author{Matteo Ciardi}
\email{matteo.ciardi@tuwien.ac.at}
\affiliation{Institute for Theoretical Physics, TU Wien, Wiedner Hauptstraße 8-10/136, 1040 Vienna, Austria}

\author{Fabio Cinti}
\email{fabio.cinti@unifi.it}
\affiliation{Dipartimento di Fisica e Astronomia, Universit\`a di Firenze, I-50019, Sesto Fiorentino (FI), Italy}
\affiliation{INFN, Sezione di Firenze, I-50019, Sesto Fiorentino (FI), Italy}

\date{\today}

\begin{abstract}
We investigate the finite-temperature phase diagram of polar molecules confined in a quasi-two-dimensional geometry by a harmonic potential along the polarization axis. We employ Quantum Monte Carlo simulations to explore the strongly correlated regime accessible with current experimental setups. By tuning temperature and confinement strength, we identify a rich set of phases, including normal fluid, superfluid, supersolid, cluster crystal, and bilayer crystal states.
Our results reveal the emergence of crystallization upon increasing temperature, highlighting the nontrivial role of thermal fluctuations in dipolar systems. In particular, we show that a bilayer crystal with one molecule per lattice site can be stabilized by varying the confinement strength at fixed interaction. Moreover, we show evidence of layering of superfluid states with phase coherence between the two layers. These findings provide insight into the interplay between interactions, confinement, and temperature in low-dimensional dipolar systems, and suggest new directions for engineering quantum phases with ultracold polar molecules.
\end{abstract}


\maketitle

\section{Introduction}
Ultracold gases have established themselves as versatile platforms for quantum simulation, enabling controlled realizations of paradigmatic models of condensed matter physics~\cite{Bloch2008, Schfer2020}. Landmark achievements include the observation of Bose–Einstein condensates (BECs)~\cite{Anderson1995, Davis1995, Proukakis2025}, self-bound droplets~\cite{Schmitt2016, Chomaz2016}, and supersolids~\cite{Bottcher2019a, Tanzi2019, Chomaz2019}, where long-range interactions give rise to exotic forms of quantum matter. 

For dipolar supersolids, theoretical investigations predict a rich ground-state phase diagram in (quasi-)two-dimensional systems showcasing triangular, honeycomb, and stripe supersolid states when the long-range interaction dominates over the short-range repulsion, and a homogeneous superfluid state in the opposite regime \cite{zhang2019, Schmidt2022, Ripley2023, Lima2025}. Theoretical approaches used to describe those states are either based on the solution of the extended Gross–Pitaevskii equation~\cite{Bisset2016, Wachtler2016, Wachtler2016_1}, with the addition of the Lee-Huang-Yang correction to the mean-field theory to avoid the collapse of the system~\cite{Lee1957, Lima2011}, or in quantum Monte carlo simulations to describe strong-correlated regimes~\cite{Jain2011, Cinti2017a, Cinti2017b, Saito2016, Macia2016, Bottcher2019b, Boninsegni2021, Kora2019}. Experimentally, supersolids have now been realized in quasi-one- and two-dimensional platforms~\cite{Sohmen2021, Norcia2021, Sinha2025, Recati2023}. 

Dipolar atoms have played a central role in these advances, but they offer limited tunability: the short-range scattering length can be controlled, while the dipolar interaction strength is fixed by the atomic magnetic moment~\cite{Chomaz2022}. As a result, only a subset of the theoretically predicted dipolar quantum phases can be accessed.
Polar molecules, by contrast, provide a high degree of control over the dipolar interaction through rotational-state dressing with static and microwave fields~\cite{Schindewolf2026, Karman2018, Anderegg2021, Schindewolf2022}.

\begin{figure}[t]
\centering
\includegraphics[width=0.5\textwidth]{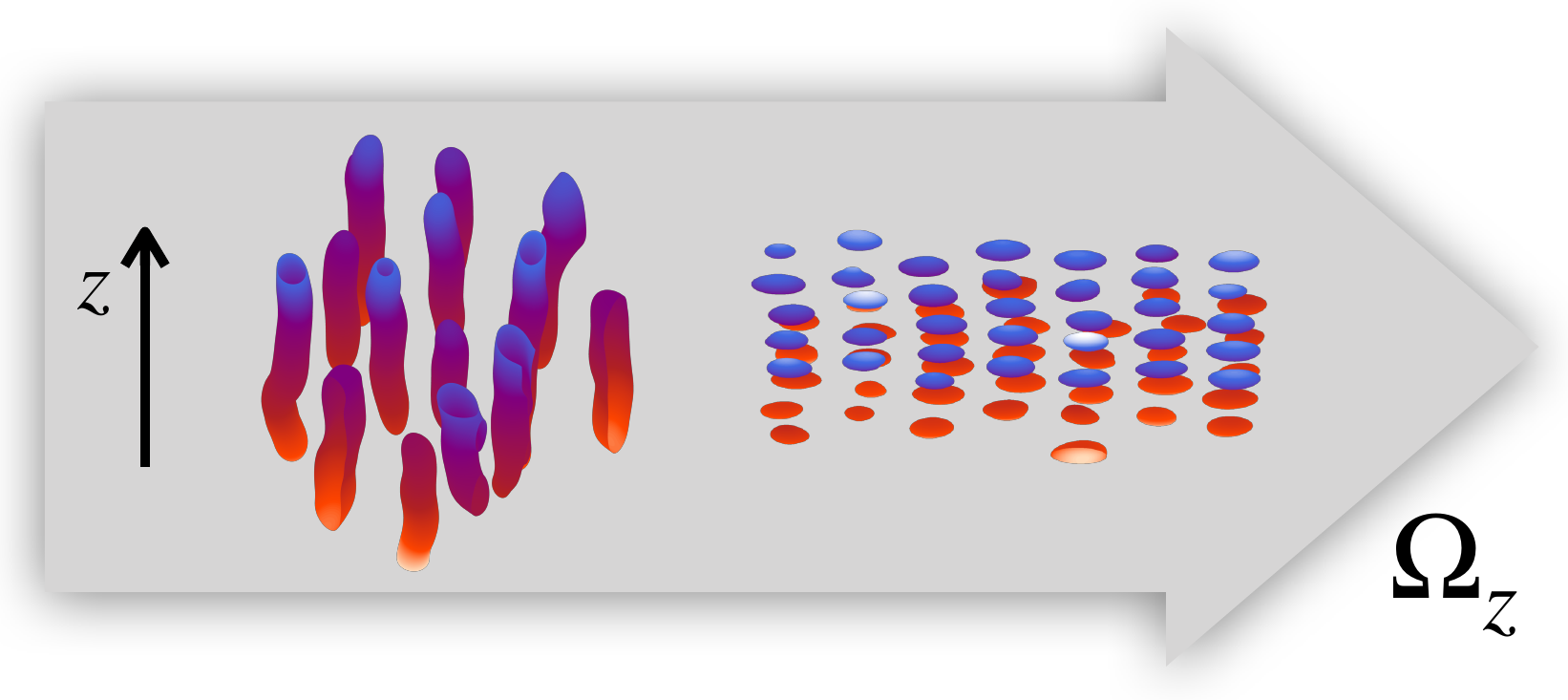}
\caption{Graphic representation of the bilayer state. In blue and red we have a snapshot of the worldlines sampled in the path-integral Monte Carlo simulation (see text). The system is made of molecules polarized along the z axis and confined by a harmonic potential of intensity $\Omega_z$ in the polarization direction (represented in  black).}
\label{fig:1}
\end{figure}

This tunability enables access to regimes ranging from cluster crystals with thousands of particles per site to crystals with a single particle per site, potentially realizing long-sought phases such as supersolids by delocalization of defects in the crystal~\cite{Andreev1971, Cinti2014}. With the recent achievement of BECs of NaCs and NaRb~\cite{Bigagli2024, Shi2025}, ultracold polar molecules now open a new avenue for quantum simulation, extending the reach of dipolar physics far beyond what is possible with atoms. The results of this experimental realization have been reproduced theoretically using the extended finite-temperature Gross-Pitaevskii equation~\cite{Baena2025} and, importantly, universality in the microwave shielding has been shown to allow the realization of BEC to be achieved with most polar molecules~\cite{Dutta2025}. Regarding the interaction potential, its derivation has been theoretically proposed years ago, and analytical expressions have been refined for both single- and double microwave fields~\cite{Buchler2007, Michelli2007, Karman2018, Lassabliere2018, Deng2025}.

 Recently, theoretical results have reported the self-assembly of dipolar droplets~\cite{Langen2025} and monolayer crystals with one molecule per site~\cite{Ciardi2025}. The effects of coherence between layers of two-dimensional (2D) ultracold gases is also a topic of interest. Previous calculations revealed coupling between superfluid layers in a system of polar molecules trapped in a 2D double-well potential~\cite{wang07}. With ultracold atoms, this coherence between two layers was also realized recently in flat geometries~\cite{Rydow2025} and theoretically in curved space~\cite{Cinti2026}. Layering is relevant for a plethora of topics in physics including ferromagnetism in multi-component quantum Hall systems~\cite{Eisenstein1996,Girvin1996}, the effects on the Berezinskii-Kosterlitz-Thouless (BKT) transition described by vortex and antivortex unbinding~\cite{Perali2013}, and exciton condensation~\cite{Gao2023}.

 \begin{figure}[t]
    \centering
    \includegraphics[width=1.0\linewidth]{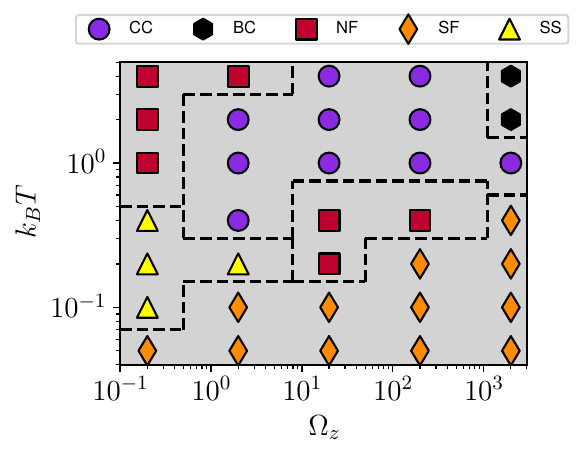}
    \caption{Phase diagram for the system of polar molecules confined in the $z$ direction by a harmonic potential $V(z) = \Omega_z z^2$ as a function of $\Omega_z$ and temperature $T$. The phase diagram displays superfluid (SF), supersolid (SS), normal fluid (NF), cluster crystal (CC), and bilayer crystal (BC) states.}
    \label{fig:2}
\end{figure}

New physics is expected to emerge in the strongly-correlated regime of polar molecules, in which the optimal theoretical approach is through Quantum Monte Carlo (QMC) simulations.

In this work, we aim to investigate the finite-temperature phase diagram of dipolar molecules under harmonic confinement along the polarization axis. By selecting interaction strengths that can be realized in quantum simulations with polar molecules, we report the observation of normal fluid (NF), supefluid (SF), supersolid (SS), cluster crystal (CC), and bilayer crystal (BC) phases (see Fig.~\ref{fig:1} for an illustration of the CC and BC states). The phase diagram obtained by tuning the confinement strength and temperature is showcased in Fig.~\ref{fig:2}. Notably, we observe crystallization with increasing temperature and demonstrate how the BC with one molecule per site can be realized by tuning the strength of the confinement with fixed interaction, which can lead to new routes towards supersolids in dipolar systems. However, in the regime of parameters considered in this work, supersolid states were observed only in cluster crystals.

The paper is organized as follows: In Section II we are presenting the model, the methodology, and the main observables. Then results are presented in Section III. 
Finally we propose a discussion of the outcomes in Section IV.

\section{System and methods}

We consider a system of $N$ polar molecules of mass $m$ polarized along the $Z$-axis.  The pairwise interaction among these molecules was devised in Ref.~\cite{Buchler2007} and reads
\begin{equation}
V(\mathbf{R}) = C_3\left(\frac{1}{R^{3}} - 3\frac{Z^2}{R^{5}}\right) + \frac{C_6}{R^{6}},
\end{equation}
with $\mathbf{R}=(X,Y,Z)$ being the relative coordinates between a pair of molecules. The $C_3$ and $C_6$ parameters depend on the dipole moment $d$  of the molecules, but also can be tuned via microwave shielding by changing the Rabi frequency $\Omega$ and the detuing $\Delta$~\cite{Deng2023}. Specifically,

\begin{equation}\label{eq:c3}
C_3 = \frac{d^2\Omega^2}{48\pi\epsilon_0(\Omega^2 + \Delta^2)},
\end{equation}
and
\begin{equation}\label{eq:c6}
C_6 = \frac{d^4\Omega^2}{128\pi^2\epsilon_0^2\hbar(\Omega^2 + \Delta^2)^{\frac{3}{2}}},
\end{equation}
with $\epsilon_0$ being the vacuum permittivity.

Choosing $\ell = (C_6/C_3)^{1/3}$ and $\epsilon = \hbar^2/(2m\ell^2)$ as the unit of length and energy, respectively, the dimensionless Hamiltonian reads:
\begin{equation}\label{eq:h}
H = -\frac{1}{2}\sum_{i=1}^N  \nabla_i^2 + \sum_{j>i=1}^N U(\mathbf{r}_j - \mathbf{r}_i) + \Omega_z\sum_{i=1}^N z_i^2,
\end{equation}
with $\mathbf{r}_i = (x_i,y_i,z_i)$ being the dimensionless coordinates of the $i$-th particle, $i = 1,2,\cdots,N$, while the second term of the Hamiltonian is $U(\mathbf{r}) = V(\mathbf{R})/\epsilon = D(r^{-3} -3z^2r^{-5}+r^{-6})$. Moreover, $D = (2m/\hbar^2)(C_3^4/C_6)^{1/3}$ is the interaction strength, and $\Omega_z = m\omega_z^2/2$ is the confinement parameter along the polarization axis, where $\omega_z$ is the harmonic potential frequency.

We investigate this system using state-of-the-art Path-Integral Monte Carlo (PIMC) simulations~\cite{Ceperley1995} with the sampling of particle permutations boosted by the worm algorithm~\cite{Boninsegni2006}. The PIMC method maps each quantum particle into classical polymers (or worldlines) through Feynman's path-integral formalism, allowing the simulation of a quantum system in a classical computer. To this aim, we consider a discretization of the imaginary-time paths of each particle in a number of $M$ time slices. The primitive approximation was used to calculate the density matrix via its convolution property; the error introduced by the discretization can always be made smaller than the statistical error of Monte Carlo methods. For the lower temperatures, we considered up to $M = 1024$ time slices. The biggest system we consider contains $N = 180$ molecules in a box of size $L_x = 9$ along the $x$ direction, $L_y = 5\sqrt{3}$ along the $y$ direction, with periodic boundary conditions in $x$ and $y$. In all cases, the averaged density in the $x-y$ plane is $\bar{\rho} =2.31$. A very large box is considered along $z$ with open boundary conditions to ensure that confinement is only due to the trapping potential. For other relevant parameters, we considered $D = 35$ in most simulations, which is experimentally feasible with NaCs molecules, $2\times10^{-1} \leq \Omega_z \leq 2\times 10^{3}$, and $5\times10^{-2}\leq k_BT \leq 4$.

To characterize crystalline phases, we define the structure factor in the $x-y$ plane,
\begin{equation}
S(\mathbf{k}) = \int d\mathbf{r_{\perp}} \rho(\mathbf{r_{\perp}}) \mathrm{e}^{-i\mathbf{k}\cdot\mathbf{r_{\perp}}},
\end{equation}
for $\mathbf{r_\perp}=(x,y)$ and $\mathbf{k}=(k_x,k_y)$. We further process our data calculating the angle-averaged structure factor $\bar{S}(k) = (2\pi)^{-1}\int d\phi S(\mathbf{k})$ and the momentum-space contrast
\begin{equation}\label{eq:qk}
q(k) = \frac{1}{2\pi}\int d\phi \left[S(\mathbf{k})-\bar{S}(k)\right]^2.
\end{equation}
By computing the contrast at the characteristic wave length $k_0$ located at the Bragg peaks of the structure factor. We locate the transition from homogeneous to modulated phases whenever $q(k_0)$ substantially increases~\cite{Ciardi2025}. We measure the phase coherence in the system by computing the superfluid fraction $f_s$ via the winding number estimator
\begin{equation}\label{eq:superfluid}
f_s = \frac{1}{2N\beta}\left(\langle |\mathbf{W}|^2\rangle - \langle\mathbf{W}\rangle^2\right)\,,
\end{equation}
where $\beta^{-1}=k_BT$, while $\mathbf{W}=(W_x,W_y,W_z)$ is the vector of winding numbers in each spatial direction~\cite{Pollock1987}, $\langle\cdots\rangle$ denotes the equilibrium average. We emphasize that $W_z = 0$ in our calculations, since the confinement forbids winding in this direction.

In the strong-interaction regime we considered in this work, the measurement of the winding numbers in the simulation may be challenging and leads to an underestimation of the statistical error associated to the  superfluid fraction $f_s$, as reported in \eqref{eq:superfluid}. In order to have a stronger evidence of phase coherence in the system we also display the frequency of permutation cycles among the molecules. 

We define this quantity as the probability, $P(L)$, that a randomly selected particle is part of a cycle of length $L$, essentially a closed loop formed by $L$ interconnected worldlines. If ${\cal N}_L(C)$ represents the number of $L$-cycles in a given configuration $C$ (where $\sum_L L{\cal N}_L(C) = N$), the probability is given by $P(L) = L\langle{\cal N}_L(C)\rangle / N$, satisfying the normalization $\sum_L P(L) = 1$. The emergence of a flattened trend in $P(L)$ serves as a signature of superfluid-like behavior within our molecules system \cite{PhysRevLett.119.215302,ciardi2025b}.

\section{Results}
\begin{figure}[t]
    \centering
    \includegraphics[width=1.0\linewidth]{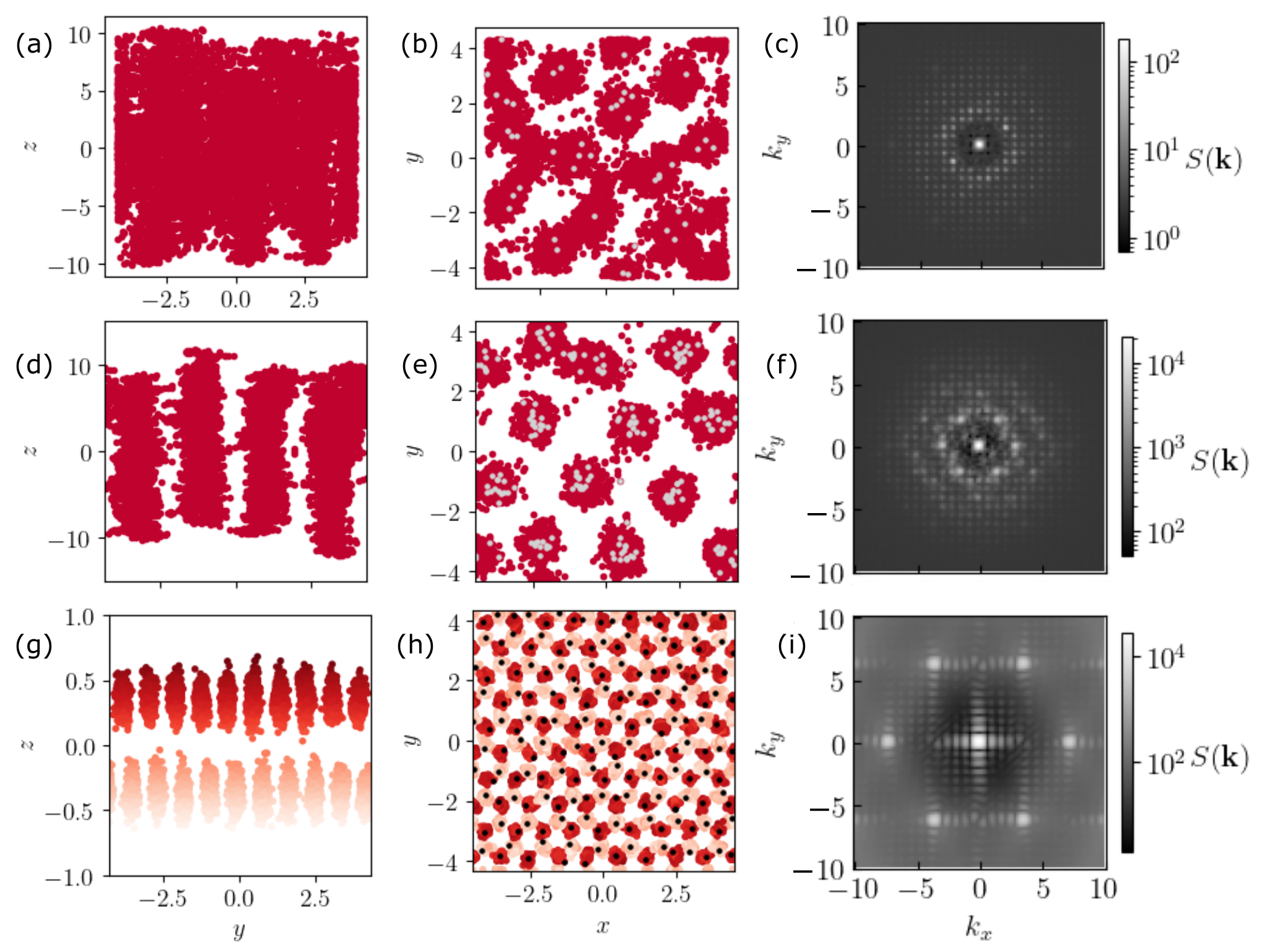}
    \caption{(a) Snapshot of the worldlines of the SS state in the $y-z$ plane. (b) A cut of the snapshot of the SS state in the $x-y$ plane showing the percolations; the gray dots represent a particular time slice. (c) The structure factor associated to (b). The physical parameters in (a)-(c) are $\Omega_z = 0.2$, $k_BT = 0.2$. (d) A snapshot of the PIMC world lines in the $y-z$ plane, showing the length of the filaments in the CC state. (e) A snapshot in the $x-y$ plane showing the triangular pattern. (f) The structure factor of the triangular pattern in the $x-y$ plane. The physical parameters in (d)-(f) are $\Omega_z = 0.2$, $k_BT = 0.4$. (g) Snapshot in the $y-z$ plane showing the separation in two layers. (h) Snapshot in the $x-y$ plane showing the bilayer triangular lattice, with the bottom one in pink and the top one in red. The black dots represent a particular time slice (we changed the color for better visualization); Note that we have one molecule per lattice site. (i) The structure factor of the density pattern in the $x-y$ plane. In (h)-(i) the physical parameters are $\Omega_z = 2000$, $k_BT = 2$, $D = 35$. In all situations we considered $D = 35$ and $\bar{\rho}=2.31$.}
    \label{fig:3}
\end{figure}

\begin{figure}[t]
\centering
\includegraphics[width=0.5\textwidth]{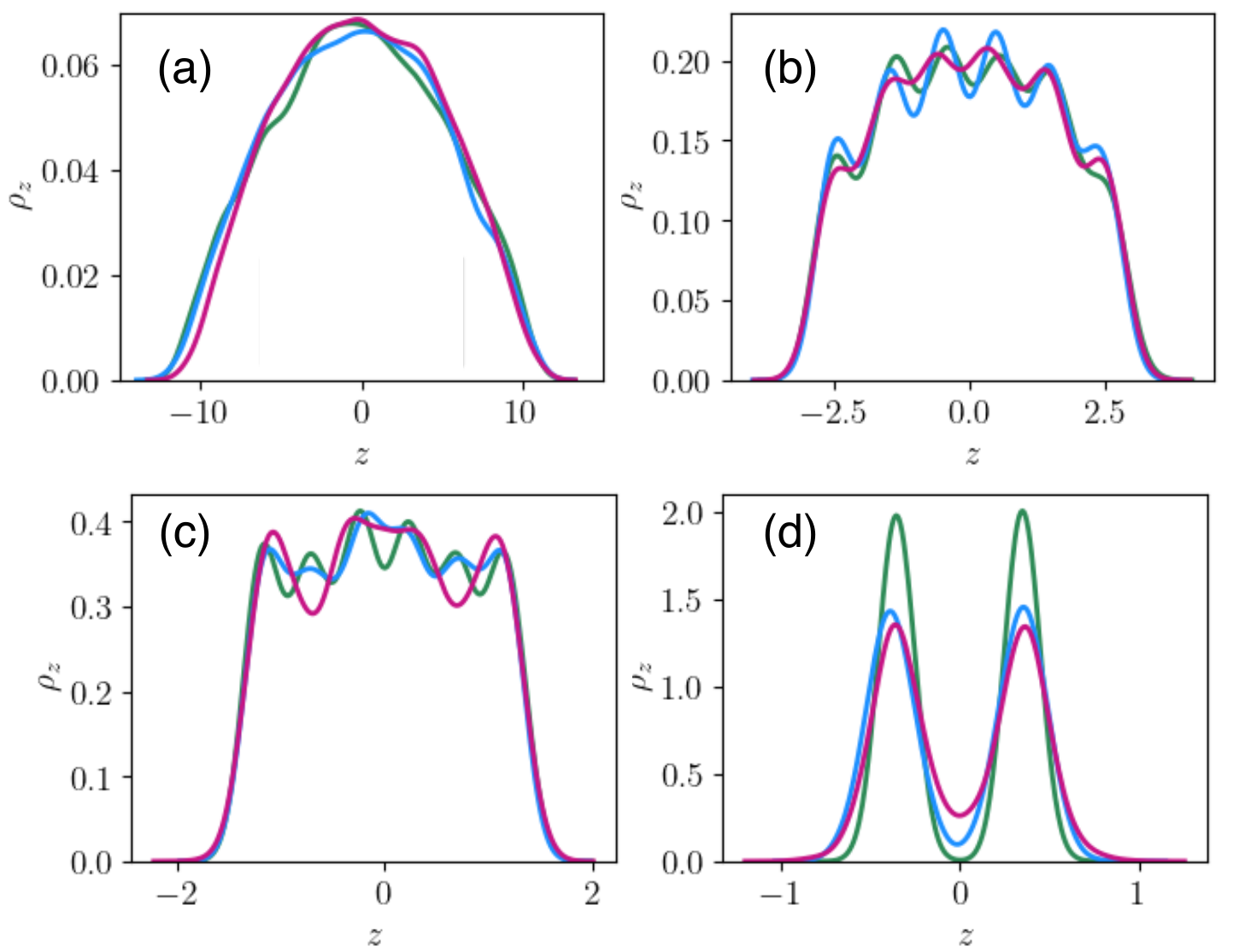}
\caption{Snapshot of particle positions along $z$ versus $\rho_z$ for different temperatures and confinement strengths ($T=2.0$ green line, $T=1.0$ light blue line, and $T=0.4$ violet line). (a) Weak confinement with $\Omega_z = 0.2$. For intermediate confinement (b) $\Omega_z = 20$ and (c) $\Omega_z = 200$ we observe density modulations that depend on the temperature. (d) Strong confinement regime $\Omega_z = 2000$, where the system splits in two layers as the temperature increases. The density of the system is $\rho=2.31$ and the interaction strength is $D = 35$.}
\label{fig:4}
\end{figure}

In Fig.~\ref{fig:2}, we show the phases of the system by varying the temperature $T$ and the confinement parameter $\Omega_z$. At the lowest temperatures, the system is in a homogeneous and SF state. As expected, when the temperature increases, the system loses its superfluid response and enters either in a NF, a CC, or a SS state. Interestingly, heating is causing crystallization in the system at fixed $\Omega_z$. In classical bilayer systems made of dipoles, such liquid to solid transition was observed by changing the interlayer spacing rather than the temperature~\cite{Lu2008}. Crystallization in dipolar bosonic gases containing two layers was also observed by tunning the system density and the distance between the layers~\cite{Cinti2017}. Interestingly, we also observe this phenomenon even in regimes where layering does not occur. Moreover, thermocrystalization from superfluid to supersolid was reported to ocuur in atomic dipolar systems due to a softenning of the roton mode~\cite{SnchezBaena2023}, an aspect also present in our system.

A depiction of the typical phases of the system is shown in Fig.~\ref{fig:3}. For weak confinement and intermediate temperature the molecules group in filaments with multiple particles. These clusters of particles present some sort of percolation and finite superfluid fraction, characterizing the SS state; see Figs.~\ref{fig:3}(a) and  (b) for the cuts in the $y-z$ and $x-y$ planes of a snapshot of the worldlines of the PIMC simulation of this state, respectively. In Fig.~\ref{fig:3}(c), we show the structure factor associated to Fig.~\ref{fig:3}(b), averaged over many PIMC samples. By increasing the confinement strength, the superfluid fraction vanishes and we observe the CC state, characterized in Figs.~\ref{fig:3}(d-f). Strikingly, at the strongest confinement, where the hard-core repulsion dominates, the system splits into two layers along the polarization direction $z$. The effect of the long-range interaction in this situation is to organize the molecules in each layer into a triangular lattice with one molecule per site. This BC state has a shift between the two layers in order to reduce the hard-core repulsion; the molecules in the upper layer are placed on the edges of the lower one (see Figs.~\ref{fig:3}(g) and (h) for the cuts of a snapshot of the BC state in the $y-z$ and $x-y$ planes, respectively, and see Fig.~\ref{fig:3}(i) for its structure factor).

The one-dimensional density $\rho_z$ along $z$ was also computed from a histogram containing the $z_i$ coordinate of each particle in each time slice of a snapshot of the simulation. Fig.~\ref{fig:4} shows $\rho_z$ for three different temperatures ($k_BT$=2.0,\,1.0,\,0.4) for different $\Omega_z$. 
Considering Fig.~\ref{fig:4}(a) ($\Omega_z=0.2$), where a weak confinement is imposed, we observe a behavior that displays a uniform character along the $z$-direction for all temperatures. Dissimilarities are noticed in panel (b). In this case, looking at $T$=2.0 and 1.0, central oscillations identify the CC phase. The oscillations soften at lower temperature, here the system depicts a NF phase, accordingly with the phase diagram in Fig.~\ref{fig:2}. Such a behavior is substantially spotted in Fig.~\ref{fig:4}(c) with $\Omega_z=200$. Finally, the last panel, Fig.~\ref{fig:4}(d), shows the system's behavior for $\Omega_z=2000$. Upon lowering the temperature, the system evolves from the BC phase into a state featuring two distinct superfluid planes. In this regime, phase coherence extends along the $z$-axis as well. Ultimately, at $T=0.4$, the system displays a clearly anisotropic superfluid character.

\begin{figure}[t]
    \centering
    \includegraphics[width=1.0\linewidth]{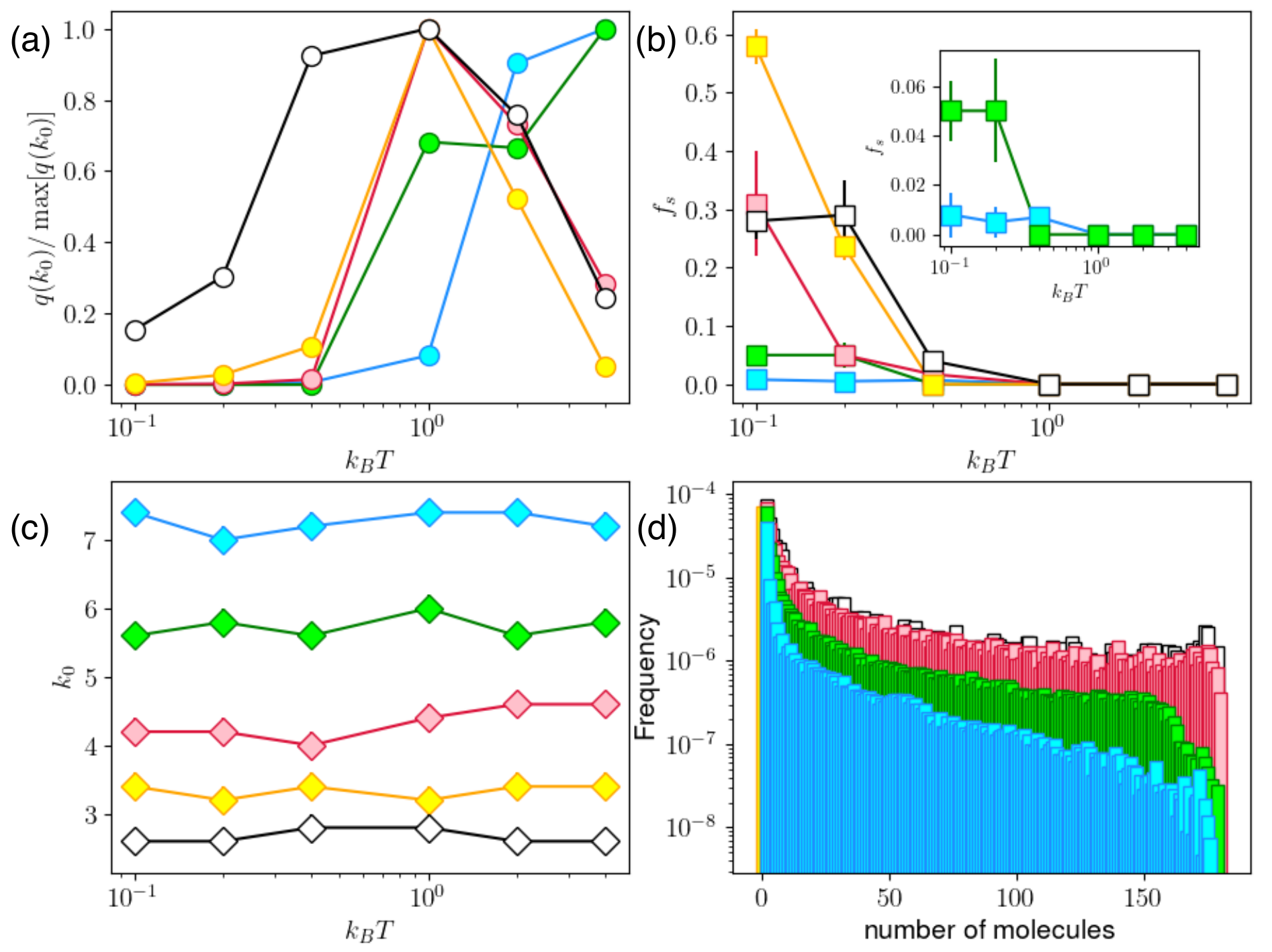}
    \caption{(a) Structure-factor contrast at the modulation wave vector (circles). (b) Superfluid fraction (squares), where the inset shows finite winding numbers at strong confinement and low temperature. (c) Characteristic momentum $k_0$ (rhombus). (d) Frequency of permutation cycles among the molecules for $k_BT = 0.2$. We consider different confinements: $\Omega_z = 0.2$ (white), $\Omega_z = 2.0$ (yellow), $\Omega_z = 20.0$ (red), $\Omega_z = 200.0$ (green), and $\Omega_z = 2000.0$ (blue). These data refer to the interaction strength $D = 35$, and average density $\bar{\rho}=2.31$. The value $\max[q(k_0)]$ is the maximum value of the contrast at fixed $\Omega_z$ considering all temperatures.}
    \label{fig:5}
\end{figure}

We further characterize the states of the system calculating the structure-factor contrast $q(k)$ in Eq.~\eqref{eq:qk}, and  the superfluid fraction $f_s$ in Eq.~\eqref{eq:superfluid}. The values of $q(k_0)$, with $k_0$ being the characteristic momentum of the crystalline structure, and $f_s$ extracted from our simulations at varying temperature $T$ and confinement $\Omega_z$ are shown in Figs.~\ref{fig:5}(a) and (b), allowing the distinction between homogeneous phases, such as the SF and the NF ones, that have vanishing contrast $q(k_0)$, and crystalline phases, like the SS, CC, and BC ones, where $q(k_0)$ is finite. A contrasting behavior behavior between weak to intermediate and strong confinement is shown in Fig.~\ref{fig:5}(a). For weak to intermediate confinement, $q(k_0)$ starts to decrease for high temperatures, indicating a softening of the density modulation in the system. In the order hand, for strong confinement, the contrast $q(k_0)$ remains strong at high temperature and in this regime we observe the formation of the BC. The differentiation between CC and BC order has been already discussed with the one-dimensional density $\rho_z$ in Fig.~\ref{fig:4}. Importantly, the number of clusters, and of particles per cluster, in the CC state depends on the confinement parameter $\Omega_z$. For increasing $\Omega_z$, the number of clusters increases and the number of particles in each cluster decreases until the formation of the BC state with one molecule per cluster. Of course, as the number of clusters increases in the crystalline phase, the lattice constant diminishes and the characteristic momentum $k_0$ increases. We estimated the values of $k_0$ from the structure factors, see Fig.~\ref{fig:3}(c) and (f) for reference, and plotted them if Fig.~\ref{fig:5}(c), where we see the dependence in $\Omega_z$, but not in the temperature $T$. The superfluid fraction $f_s$ is used to distinguish the SF and SS phases from the NF, CC, and BC ones. As mentioned, the computation of $f_s$ in strong-correlated dipolar systems is challenging, therefore we also analyze the length of molecular permutation cycles to characterize superfluidity, see Fig.~\ref{fig:5}(d). A plateau of long permutation cycles is indicative of phase coherence and therefore superfluid behavior. For instance, at $k_BT = 0.4$ and $\Omega_z = 2000$, we display a small but finite value of $f_s$ in the inset of Fig.~\ref{fig:5}(b) which is accompanied by long permutation cycles, see Fig.~\ref{fig:5}(d), thus characterizing one of the SF states found in this system. Strikingly, this state has strong modlutations in $\rho_z$, as displayed in Fig.~\ref{fig:4}(d), therefore representing coherence between layered superfluids.

Now, we investigate the phases of the system in different interaction regimes. In fig.~\ref{fig:a1}(a)-(c), we display snapshots of the PIMC simulation for $D=127$, $\Omega_z = 200$, and $k_BT = 2$, where we see clusterization and hints of multiple layering. Interestingly, we also observed percolation between some of the clusters as in the SS state represented in Figs.~\ref{fig:3}(b). We also considered a weaker interaction $D=11$ in fig.~\ref{fig:a1}(d)-(e) maintaining the confinement strength and temperature. In this situation the percolation is more evident, but we still notice a tendency of layering formation. At this regime of temperature, the has no superfluid fraction.

\section{Discussion}

We performed Quantum Monte Carlo simulations of a system made of bosonic polar molecules confined by a harmonic potential along the polarization axis considering periodic boundary conditions in the plane perpendicular to the former axis. By varying the system temperature $T$ and the confinement strength $\Omega_z$ we construct a phase diagram for this system at a fixed interaction $D = 35$. We observe a homogeneous SF phase that, with increasing temperature, transitions to either a homogeneous NF, a modulated CC, or a SS that breaks simultaneously translational and gauge symmetries.
Such an interesting thermo-crystallization effect in dipolar gases has been experimentally observed recently in Ref.~\cite{SnchezBaena2023} by using a cigar-shaped-like confinements.
We also observe a transition from the CC state with multiple particles per cluster to a BC with a single molecule per site. The BC state splits in two triangular lattices separated along the polarization direction. Layered quantum systems are a subject of current interest in many scientific fields ranging from quantum Hall systems to the BKT theory in coupled 2D systems.For strong confinement and low temperature, our results indicate the formation of layered SF states with particle permutation among the two layers.

\begin{figure}[t]
    \centering
    \includegraphics[width=1.0\linewidth]{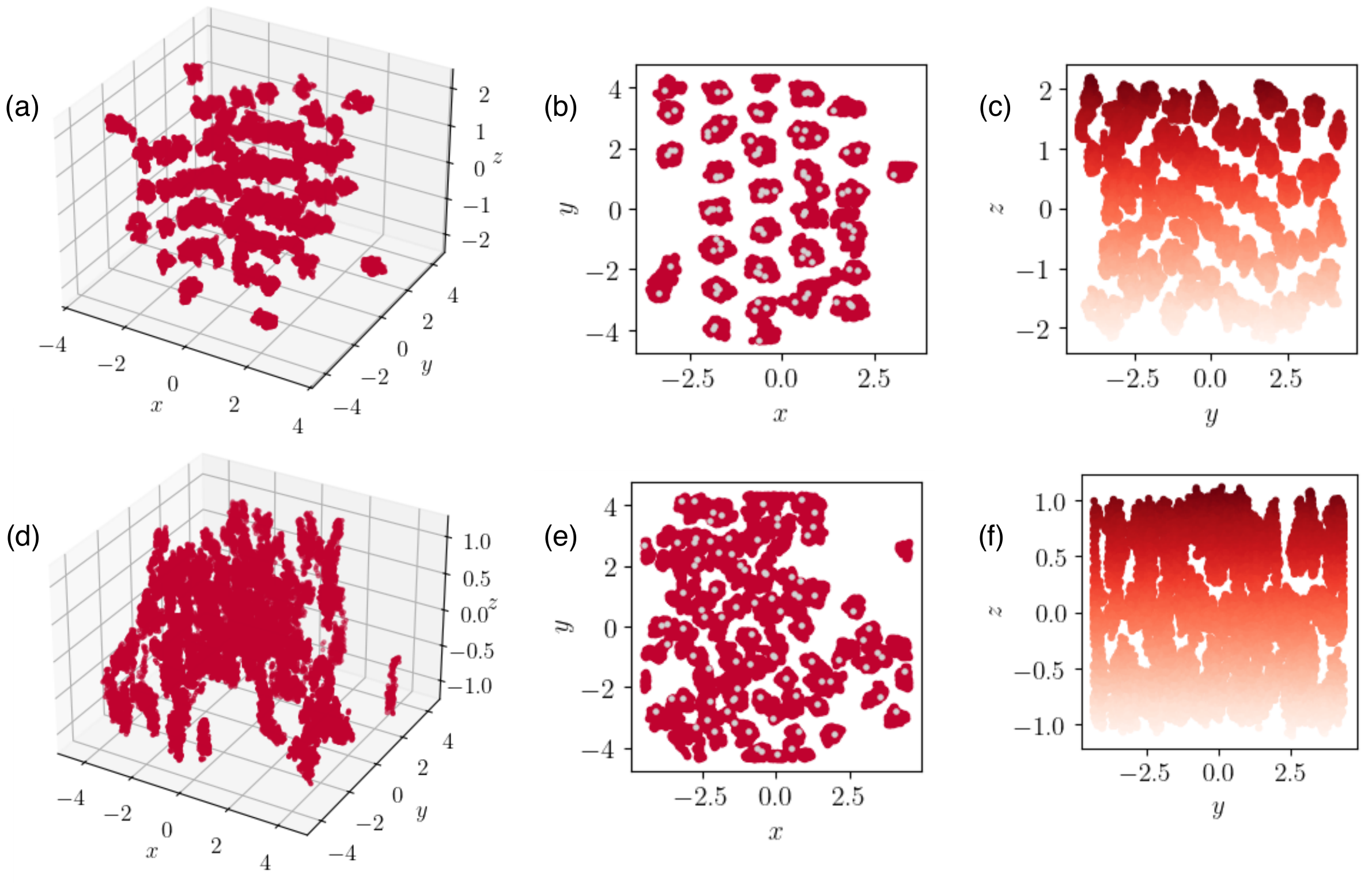}
    \caption{Snapshots of the worldlines in the PIMC simulation. In (a)-(c) the interaction strength is $D=127$ and in (d)-(f) $D = 11$. (a) and (d) depicts the three-dimensional system. Projections of the worldlines in the $x-y$ and $y-z$ planes are represented in (b,e) and (c,f), respectively. The harmonic confinement strength is $\Omega_z = 200$ and the temperature is $k_BT = 2.$}
    \label{fig:a1}
\end{figure}

Moreover, we present realistic values for our dimensionless parameters considering the sodium-cesium (NaCs) molecule. With a strong dipole moment $d = 4.7$ D, and a Rabi frequency $\Omega = 2\pi\times10$ MHz, we achieve the interaction strength $D = 35$ with a microwave detuning $\Delta \approx 0.84 \Omega$. In this regime our unit of length is $\ell \approx 73$ nm, and the system temperature ranges from $T \approx 30$ nK to $T\approx 2$ $\mu$K. All these values are experimentally feasible. Considering he trapping frequency, it lies between $\omega_z \approx 50$ kHz and $\omega_z \approx 5$ MHz, which would certainly be experimentally challenging. Considering molecules with higher dipolar moment, as the  strontium oxide (SrO) with $d = 8.9$ D, we can reach dimensionless interactions strengths up to $D = 127$, where we observed cluster formation without a crystalline (periodic) structure and hints of layer formation at weaker trapping, see Fig.~\ref{fig:a1}.

At large confinement, we observe that the cluster phases shrink, leading to an almost immediate transition from fluid to a bilayer crystal as temperature is increased. While this phase is still challenging to realize experimentally, due to the strong confinement required, we expect it to be an interesting setting to investigate the response to defects, and inter-layer correlations, in a self-assembled setting, providing a strong candidate for new forms of supersolidity.

Going further, complex confinement by external potentials as in quasicrystalline optical lattices seems to be an interesting candidate for the development of new physics as reported in Ref.~\cite{Zampronio2024}.  The aperiodic nature of the quasicrystal breaks Bloch's theorem, a cornerstone of solid state physics, thus raising a fundamental question. Fractal structures not only show aperiodicity but can possess noninteger dimension, which could reveal new mechanisms of superfluidity. Quasicrystal~\cite{Yu2024} and fractal~\cite{Kempkes2018, Verstraten2025} confinements are subject of recent interest in quantum simulation. Nonetheless, the behavior of strong-correlated bosons that experiment long-range interaction, as can be realized in polar-molecules platforms, is still elusive and could possibily reveal new routes towards supersolidity. In future simulations, we will explore how the lattice geometry, imposed by an external potential, affects the behavior of cold polar molecules.

\begin{acknowledgments}
F. C. and V. Z. aknowledge finantial suport from PNRR MUR Project No. PE0000023-NQSTI. V. Z. also aknowledges finantial support from Fundação de Amparo à Pesquisa do Estado de São Paulo (Fapesp). M.~C. acknowledges funding from the Austrian Science Fund (Grant No. 10.55776/COE1). We thank the NICIS Centre for High-Performance Computing, South Africa, and the Centro Nacional de Processamento de Alto Desempenho em São Paulo (CENAPAD), Brazil, for providing computational resources.
\end{acknowledgments}

\bibliography{dip.bib}

\begin{thebibliography}{66}%
\makeatletter
\providecommand \@ifxundefined [1]{%
 \@ifx{#1\undefined}
}%
\providecommand \@ifnum [1]{%
 \ifnum #1\expandafter \@firstoftwo
 \else \expandafter \@secondoftwo
 \fi
}%
\providecommand \@ifx [1]{%
 \ifx #1\expandafter \@firstoftwo
 \else \expandafter \@secondoftwo
 \fi
}%
\providecommand \natexlab [1]{#1}%
\providecommand \enquote  [1]{``#1''}%
\providecommand \bibnamefont  [1]{#1}%
\providecommand \bibfnamefont [1]{#1}%
\providecommand \citenamefont [1]{#1}%
\providecommand \href@noop [0]{\@secondoftwo}%
\providecommand \href [0]{\begingroup \@sanitize@url \@href}%
\providecommand \@href[1]{\@@startlink{#1}\@@href}%
\providecommand \@@href[1]{\endgroup#1\@@endlink}%
\providecommand \@sanitize@url [0]{\catcode `\\12\catcode `\$12\catcode `\&12\catcode `\#12\catcode `\^12\catcode `\_12\catcode `\%12\relax}%
\providecommand \@@startlink[1]{}%
\providecommand \@@endlink[0]{}%
\providecommand \url  [0]{\begingroup\@sanitize@url \@url }%
\providecommand \@url [1]{\endgroup\@href {#1}{\urlprefix }}%
\providecommand \urlprefix  [0]{URL }%
\providecommand \Eprint [0]{\href }%
\providecommand \doibase [0]{https://doi.org/}%
\providecommand \selectlanguage [0]{\@gobble}%
\providecommand \bibinfo  [0]{\@secondoftwo}%
\providecommand \bibfield  [0]{\@secondoftwo}%
\providecommand \translation [1]{[#1]}%
\providecommand \BibitemOpen [0]{}%
\providecommand \bibitemStop [0]{}%
\providecommand \bibitemNoStop [0]{.\EOS\space}%
\providecommand \EOS [0]{\spacefactor3000\relax}%
\providecommand \BibitemShut  [1]{\csname bibitem#1\endcsname}%
\let\auto@bib@innerbib\@empty
\bibitem [{\citenamefont {Bloch}\ \emph {et~al.}(2008)\citenamefont {Bloch}, \citenamefont {Dalibard},\ and\ \citenamefont {Zwerger}}]{Bloch2008}%
  \BibitemOpen
  \bibfield  {author} {\bibinfo {author} {\bibfnamefont {I.}~\bibnamefont {Bloch}}, \bibinfo {author} {\bibfnamefont {J.}~\bibnamefont {Dalibard}},\ and\ \bibinfo {author} {\bibfnamefont {W.}~\bibnamefont {Zwerger}},\ }\bibfield  {title} {\bibinfo {title} {Many-body physics with ultracold gases},\ }\href {https://doi.org/10.1103/RevModPhys.80.885} {\bibfield  {journal} {\bibinfo  {journal} {Rev. Mod. Phys.}\ }\textbf {\bibinfo {volume} {80}},\ \bibinfo {pages} {885} (\bibinfo {year} {2008})}\BibitemShut {NoStop}%
\bibitem [{\citenamefont {Sch\"{a}fer}\ \emph {et~al.}(2020)\citenamefont {Sch\"{a}fer}, \citenamefont {Fukuhara}, \citenamefont {Sugawa}, \citenamefont {Takasu},\ and\ \citenamefont {Takahashi}}]{Schfer2020}%
  \BibitemOpen
  \bibfield  {author} {\bibinfo {author} {\bibfnamefont {F.}~\bibnamefont {Sch\"{a}fer}}, \bibinfo {author} {\bibfnamefont {T.}~\bibnamefont {Fukuhara}}, \bibinfo {author} {\bibfnamefont {S.}~\bibnamefont {Sugawa}}, \bibinfo {author} {\bibfnamefont {Y.}~\bibnamefont {Takasu}},\ and\ \bibinfo {author} {\bibfnamefont {Y.}~\bibnamefont {Takahashi}},\ }\bibfield  {title} {\bibinfo {title} {Tools for quantum simulation with ultracold atoms in optical lattices},\ }\href {https://doi.org/10.1038/s42254-020-0195-3} {\bibfield  {journal} {\bibinfo  {journal} {Nature Reviews Physics}\ }\textbf {\bibinfo {volume} {2}},\ \bibinfo {pages} {411–425} (\bibinfo {year} {2020})}\BibitemShut {NoStop}%
\bibitem [{\citenamefont {Anderson}\ \emph {et~al.}(1995)\citenamefont {Anderson}, \citenamefont {Ensher}, \citenamefont {Matthews}, \citenamefont {Wieman},\ and\ \citenamefont {Cornell}}]{Anderson1995}%
  \BibitemOpen
  \bibfield  {author} {\bibinfo {author} {\bibfnamefont {M.~H.}\ \bibnamefont {Anderson}}, \bibinfo {author} {\bibfnamefont {J.~R.}\ \bibnamefont {Ensher}}, \bibinfo {author} {\bibfnamefont {M.~R.}\ \bibnamefont {Matthews}}, \bibinfo {author} {\bibfnamefont {C.~E.}\ \bibnamefont {Wieman}},\ and\ \bibinfo {author} {\bibfnamefont {E.~A.}\ \bibnamefont {Cornell}},\ }\bibfield  {title} {\bibinfo {title} {Observation of bose-einstein condensation in a dilute atomic vapor},\ }\href {https://doi.org/10.1126/science.269.5221.198} {\bibfield  {journal} {\bibinfo  {journal} {Science}\ }\textbf {\bibinfo {volume} {269}},\ \bibinfo {pages} {198} (\bibinfo {year} {1995})}\BibitemShut {NoStop}%
\bibitem [{\citenamefont {Davis}\ \emph {et~al.}(1995)\citenamefont {Davis}, \citenamefont {Mewes}, \citenamefont {Andrews}, \citenamefont {van Druten}, \citenamefont {Durfee}, \citenamefont {Kurn},\ and\ \citenamefont {Ketterle}}]{Davis1995}%
  \BibitemOpen
  \bibfield  {author} {\bibinfo {author} {\bibfnamefont {K.~B.}\ \bibnamefont {Davis}}, \bibinfo {author} {\bibfnamefont {M.~O.}\ \bibnamefont {Mewes}}, \bibinfo {author} {\bibfnamefont {M.~R.}\ \bibnamefont {Andrews}}, \bibinfo {author} {\bibfnamefont {N.~J.}\ \bibnamefont {van Druten}}, \bibinfo {author} {\bibfnamefont {D.~S.}\ \bibnamefont {Durfee}}, \bibinfo {author} {\bibfnamefont {D.~M.}\ \bibnamefont {Kurn}},\ and\ \bibinfo {author} {\bibfnamefont {W.}~\bibnamefont {Ketterle}},\ }\bibfield  {title} {\bibinfo {title} {Bose-einstein condensation in a gas of sodium atoms},\ }\href {https://doi.org/10.1103/PhysRevLett.75.3969} {\bibfield  {journal} {\bibinfo  {journal} {Phys. Rev. Lett.}\ }\textbf {\bibinfo {volume} {75}},\ \bibinfo {pages} {3969} (\bibinfo {year} {1995})}\BibitemShut {NoStop}%
\bibitem [{\citenamefont {Proukakis}(2025)}]{Proukakis2025}%
  \BibitemOpen
  \bibfield  {author} {\bibinfo {author} {\bibfnamefont {N.~P.}\ \bibnamefont {Proukakis}},\ }\bibfield  {title} {\bibinfo {title} {A century of bose-einstein condensation},\ }\bibfield  {journal} {\bibinfo  {journal} {Communications Physics}\ }\textbf {\bibinfo {volume} {8}},\ \href {https://doi.org/10.1038/s42005-025-02195-x} {10.1038/s42005-025-02195-x} (\bibinfo {year} {2025})\BibitemShut {NoStop}%
\bibitem [{\citenamefont {Schmitt}\ \emph {et~al.}(2016)\citenamefont {Schmitt}, \citenamefont {Wenzel}, \citenamefont {B{\"o}ttcher}, \citenamefont {Ferrier-Barbut},\ and\ \citenamefont {Pfau}}]{Schmitt2016}%
  \BibitemOpen
  \bibfield  {author} {\bibinfo {author} {\bibfnamefont {M.}~\bibnamefont {Schmitt}}, \bibinfo {author} {\bibfnamefont {M.}~\bibnamefont {Wenzel}}, \bibinfo {author} {\bibfnamefont {F.}~\bibnamefont {B{\"o}ttcher}}, \bibinfo {author} {\bibfnamefont {I.}~\bibnamefont {Ferrier-Barbut}},\ and\ \bibinfo {author} {\bibfnamefont {T.}~\bibnamefont {Pfau}},\ }\bibfield  {title} {\bibinfo {title} {Self-bound droplets of a dilute magnetic quantum liquid},\ }\href {https://doi.org/10.1038/nature20126} {\bibfield  {journal} {\bibinfo  {journal} {Nature}\ }\textbf {\bibinfo {volume} {539}},\ \bibinfo {pages} {259} (\bibinfo {year} {2016})}\BibitemShut {NoStop}%
\bibitem [{\citenamefont {Chomaz}\ \emph {et~al.}(2016)\citenamefont {Chomaz}, \citenamefont {Baier}, \citenamefont {Petter}, \citenamefont {Mark}, \citenamefont {W\"achtler}, \citenamefont {Santos},\ and\ \citenamefont {Ferlaino}}]{Chomaz2016}%
  \BibitemOpen
  \bibfield  {author} {\bibinfo {author} {\bibfnamefont {L.}~\bibnamefont {Chomaz}}, \bibinfo {author} {\bibfnamefont {S.}~\bibnamefont {Baier}}, \bibinfo {author} {\bibfnamefont {D.}~\bibnamefont {Petter}}, \bibinfo {author} {\bibfnamefont {M.~J.}\ \bibnamefont {Mark}}, \bibinfo {author} {\bibfnamefont {F.}~\bibnamefont {W\"achtler}}, \bibinfo {author} {\bibfnamefont {L.}~\bibnamefont {Santos}},\ and\ \bibinfo {author} {\bibfnamefont {F.}~\bibnamefont {Ferlaino}},\ }\bibfield  {title} {\bibinfo {title} {Quantum-fluctuation-driven crossover from a dilute bose-einstein condensate to a macrodroplet in a dipolar quantum fluid},\ }\href {https://doi.org/10.1103/PhysRevX.6.041039} {\bibfield  {journal} {\bibinfo  {journal} {Phys. Rev. X}\ }\textbf {\bibinfo {volume} {6}},\ \bibinfo {pages} {041039} (\bibinfo {year} {2016})}\BibitemShut {NoStop}%
\bibitem [{\citenamefont {B\"ottcher}\ \emph {et~al.}(2019{\natexlab{a}})\citenamefont {B\"ottcher}, \citenamefont {Schmidt}, \citenamefont {Wenzel}, \citenamefont {Hertkorn}, \citenamefont {Guo}, \citenamefont {Langen},\ and\ \citenamefont {Pfau}}]{Bottcher2019a}%
  \BibitemOpen
  \bibfield  {author} {\bibinfo {author} {\bibfnamefont {F.}~\bibnamefont {B\"ottcher}}, \bibinfo {author} {\bibfnamefont {J.}~\bibnamefont {Schmidt}}, \bibinfo {author} {\bibfnamefont {M.}~\bibnamefont {Wenzel}}, \bibinfo {author} {\bibfnamefont {J.}~\bibnamefont {Hertkorn}}, \bibinfo {author} {\bibfnamefont {M.}~\bibnamefont {Guo}}, \bibinfo {author} {\bibfnamefont {T.}~\bibnamefont {Langen}},\ and\ \bibinfo {author} {\bibfnamefont {T.}~\bibnamefont {Pfau}},\ }\bibfield  {title} {\bibinfo {title} {Transient supersolid properties in an array of dipolar quantum droplets},\ }\href {https://doi.org/10.1103/PhysRevX.9.011051} {\bibfield  {journal} {\bibinfo  {journal} {Phys. Rev. X}\ }\textbf {\bibinfo {volume} {9}},\ \bibinfo {pages} {011051} (\bibinfo {year} {2019}{\natexlab{a}})}\BibitemShut {NoStop}%
\bibitem [{\citenamefont {Tanzi}\ \emph {et~al.}(2019)\citenamefont {Tanzi}, \citenamefont {Lucioni}, \citenamefont {Fam\`a}, \citenamefont {Catani}, \citenamefont {Fioretti}, \citenamefont {Gabbanini}, \citenamefont {Bisset}, \citenamefont {Santos},\ and\ \citenamefont {Modugno}}]{Tanzi2019}%
  \BibitemOpen
  \bibfield  {author} {\bibinfo {author} {\bibfnamefont {L.}~\bibnamefont {Tanzi}}, \bibinfo {author} {\bibfnamefont {E.}~\bibnamefont {Lucioni}}, \bibinfo {author} {\bibfnamefont {F.}~\bibnamefont {Fam\`a}}, \bibinfo {author} {\bibfnamefont {J.}~\bibnamefont {Catani}}, \bibinfo {author} {\bibfnamefont {A.}~\bibnamefont {Fioretti}}, \bibinfo {author} {\bibfnamefont {C.}~\bibnamefont {Gabbanini}}, \bibinfo {author} {\bibfnamefont {R.~N.}\ \bibnamefont {Bisset}}, \bibinfo {author} {\bibfnamefont {L.}~\bibnamefont {Santos}},\ and\ \bibinfo {author} {\bibfnamefont {G.}~\bibnamefont {Modugno}},\ }\bibfield  {title} {\bibinfo {title} {Observation of a dipolar quantum gas with metastable supersolid properties},\ }\href {https://doi.org/10.1103/PhysRevLett.122.130405} {\bibfield  {journal} {\bibinfo  {journal} {Phys. Rev. Lett.}\ }\textbf {\bibinfo {volume} {122}},\ \bibinfo {pages} {130405} (\bibinfo {year} {2019})}\BibitemShut {NoStop}%
\bibitem [{\citenamefont {Chomaz}\ \emph {et~al.}(2019)\citenamefont {Chomaz}, \citenamefont {Petter}, \citenamefont {Ilzh\"ofer}, \citenamefont {Natale}, \citenamefont {Trautmann}, \citenamefont {Politi}, \citenamefont {Durastante}, \citenamefont {van Bijnen}, \citenamefont {Patscheider}, \citenamefont {Sohmen}, \citenamefont {Mark},\ and\ \citenamefont {Ferlaino}}]{Chomaz2019}%
  \BibitemOpen
  \bibfield  {author} {\bibinfo {author} {\bibfnamefont {L.}~\bibnamefont {Chomaz}}, \bibinfo {author} {\bibfnamefont {D.}~\bibnamefont {Petter}}, \bibinfo {author} {\bibfnamefont {P.}~\bibnamefont {Ilzh\"ofer}}, \bibinfo {author} {\bibfnamefont {G.}~\bibnamefont {Natale}}, \bibinfo {author} {\bibfnamefont {A.}~\bibnamefont {Trautmann}}, \bibinfo {author} {\bibfnamefont {C.}~\bibnamefont {Politi}}, \bibinfo {author} {\bibfnamefont {G.}~\bibnamefont {Durastante}}, \bibinfo {author} {\bibfnamefont {R.~M.~W.}\ \bibnamefont {van Bijnen}}, \bibinfo {author} {\bibfnamefont {A.}~\bibnamefont {Patscheider}}, \bibinfo {author} {\bibfnamefont {M.}~\bibnamefont {Sohmen}}, \bibinfo {author} {\bibfnamefont {M.~J.}\ \bibnamefont {Mark}},\ and\ \bibinfo {author} {\bibfnamefont {F.}~\bibnamefont {Ferlaino}},\ }\bibfield  {title} {\bibinfo {title} {Long-lived and transient supersolid behaviors in dipolar quantum gases},\ }\href {https://doi.org/10.1103/PhysRevX.9.021012} {\bibfield  {journal} {\bibinfo  {journal} {Phys. Rev.
  X}\ }\textbf {\bibinfo {volume} {9}},\ \bibinfo {pages} {021012} (\bibinfo {year} {2019})}\BibitemShut {NoStop}%
\bibitem [{\citenamefont {Zhang}\ \emph {et~al.}(2019)\citenamefont {Zhang}, \citenamefont {Maucher},\ and\ \citenamefont {Pohl}}]{zhang2019}%
  \BibitemOpen
  \bibfield  {author} {\bibinfo {author} {\bibfnamefont {Y.-C.}\ \bibnamefont {Zhang}}, \bibinfo {author} {\bibfnamefont {F.}~\bibnamefont {Maucher}},\ and\ \bibinfo {author} {\bibfnamefont {T.}~\bibnamefont {Pohl}},\ }\bibfield  {title} {\bibinfo {title} {Supersolidity around a critical point in dipolar bose-einstein condensates},\ }\href {https://doi.org/10.1103/PhysRevLett.123.015301} {\bibfield  {journal} {\bibinfo  {journal} {Phys. Rev. Lett.}\ }\textbf {\bibinfo {volume} {123}},\ \bibinfo {pages} {015301} (\bibinfo {year} {2019})}\BibitemShut {NoStop}%
\bibitem [{\citenamefont {Schmidt}\ \emph {et~al.}(2022)\citenamefont {Schmidt}, \citenamefont {Lassabli\`ere}, \citenamefont {Qu\'em\'ener},\ and\ \citenamefont {Langen}}]{Schmidt2022}%
  \BibitemOpen
  \bibfield  {author} {\bibinfo {author} {\bibfnamefont {M.}~\bibnamefont {Schmidt}}, \bibinfo {author} {\bibfnamefont {L.}~\bibnamefont {Lassabli\`ere}}, \bibinfo {author} {\bibfnamefont {G.}~\bibnamefont {Qu\'em\'ener}},\ and\ \bibinfo {author} {\bibfnamefont {T.}~\bibnamefont {Langen}},\ }\bibfield  {title} {\bibinfo {title} {Self-bound dipolar droplets and supersolids in molecular bose-einstein condensates},\ }\href {https://doi.org/10.1103/PhysRevResearch.4.013235} {\bibfield  {journal} {\bibinfo  {journal} {Phys. Rev. Res.}\ }\textbf {\bibinfo {volume} {4}},\ \bibinfo {pages} {013235} (\bibinfo {year} {2022})}\BibitemShut {NoStop}%
\bibitem [{\citenamefont {Ripley}\ \emph {et~al.}(2023)\citenamefont {Ripley}, \citenamefont {Baillie},\ and\ \citenamefont {Blakie}}]{Ripley2023}%
  \BibitemOpen
  \bibfield  {author} {\bibinfo {author} {\bibfnamefont {B.~T.~E.}\ \bibnamefont {Ripley}}, \bibinfo {author} {\bibfnamefont {D.}~\bibnamefont {Baillie}},\ and\ \bibinfo {author} {\bibfnamefont {P.~B.}\ \bibnamefont {Blakie}},\ }\bibfield  {title} {\bibinfo {title} {Two-dimensional supersolidity in a planar dipolar bose gas},\ }\href {https://doi.org/10.1103/PhysRevA.108.053321} {\bibfield  {journal} {\bibinfo  {journal} {Phys. Rev. A}\ }\textbf {\bibinfo {volume} {108}},\ \bibinfo {pages} {053321} (\bibinfo {year} {2023})}\BibitemShut {NoStop}%
\bibitem [{\citenamefont {Lima}\ \emph {et~al.}(2025)\citenamefont {Lima}, \citenamefont {Grossklags}, \citenamefont {Zampronio}, \citenamefont {Cinti},\ and\ \citenamefont {Mendoza-Coto}}]{Lima2025}%
  \BibitemOpen
  \bibfield  {author} {\bibinfo {author} {\bibfnamefont {D.}~\bibnamefont {Lima}}, \bibinfo {author} {\bibfnamefont {M.}~\bibnamefont {Grossklags}}, \bibinfo {author} {\bibfnamefont {V.}~\bibnamefont {Zampronio}}, \bibinfo {author} {\bibfnamefont {F.}~\bibnamefont {Cinti}},\ and\ \bibinfo {author} {\bibfnamefont {A.}~\bibnamefont {Mendoza-Coto}},\ }\bibfield  {title} {\bibinfo {title} {Supersolid dipolar phases in planar geometry: Effects of tilted polarization},\ }\href {https://doi.org/10.1103/mlfn-114m} {\bibfield  {journal} {\bibinfo  {journal} {Phys. Rev. A}\ }\textbf {\bibinfo {volume} {111}},\ \bibinfo {pages} {063311} (\bibinfo {year} {2025})}\BibitemShut {NoStop}%
\bibitem [{\citenamefont {Bisset}\ \emph {et~al.}(2016)\citenamefont {Bisset}, \citenamefont {Wilson}, \citenamefont {Baillie},\ and\ \citenamefont {Blakie}}]{Bisset2016}%
  \BibitemOpen
  \bibfield  {author} {\bibinfo {author} {\bibfnamefont {R.~N.}\ \bibnamefont {Bisset}}, \bibinfo {author} {\bibfnamefont {R.~M.}\ \bibnamefont {Wilson}}, \bibinfo {author} {\bibfnamefont {D.}~\bibnamefont {Baillie}},\ and\ \bibinfo {author} {\bibfnamefont {P.~B.}\ \bibnamefont {Blakie}},\ }\bibfield  {title} {\bibinfo {title} {Ground-state phase diagram of a dipolar condensate with quantum fluctuations},\ }\href {https://doi.org/10.1103/PhysRevA.94.033619} {\bibfield  {journal} {\bibinfo  {journal} {Phys. Rev. A}\ }\textbf {\bibinfo {volume} {94}},\ \bibinfo {pages} {033619} (\bibinfo {year} {2016})}\BibitemShut {NoStop}%
\bibitem [{\citenamefont {W\"achtler}\ and\ \citenamefont {Santos}(2016{\natexlab{a}})}]{Wachtler2016}%
  \BibitemOpen
  \bibfield  {author} {\bibinfo {author} {\bibfnamefont {F.}~\bibnamefont {W\"achtler}}\ and\ \bibinfo {author} {\bibfnamefont {L.}~\bibnamefont {Santos}},\ }\bibfield  {title} {\bibinfo {title} {Quantum filaments in dipolar bose-einstein condensates},\ }\href {https://doi.org/10.1103/PhysRevA.93.061603} {\bibfield  {journal} {\bibinfo  {journal} {Phys. Rev. A}\ }\textbf {\bibinfo {volume} {93}},\ \bibinfo {pages} {061603(R)} (\bibinfo {year} {2016}{\natexlab{a}})}\BibitemShut {NoStop}%
\bibitem [{\citenamefont {W\"achtler}\ and\ \citenamefont {Santos}(2016{\natexlab{b}})}]{Wachtler2016_1}%
  \BibitemOpen
  \bibfield  {author} {\bibinfo {author} {\bibfnamefont {F.}~\bibnamefont {W\"achtler}}\ and\ \bibinfo {author} {\bibfnamefont {L.}~\bibnamefont {Santos}},\ }\bibfield  {title} {\bibinfo {title} {Ground-state properties and elementary excitations of quantum droplets in dipolar bose-einstein condensates},\ }\href {https://doi.org/10.1103/PhysRevA.94.043618} {\bibfield  {journal} {\bibinfo  {journal} {Phys. Rev. A}\ }\textbf {\bibinfo {volume} {94}},\ \bibinfo {pages} {043618} (\bibinfo {year} {2016}{\natexlab{b}})}\BibitemShut {NoStop}%
\bibitem [{\citenamefont {Lee}\ \emph {et~al.}(1957)\citenamefont {Lee}, \citenamefont {Huang},\ and\ \citenamefont {Yang}}]{Lee1957}%
  \BibitemOpen
  \bibfield  {author} {\bibinfo {author} {\bibfnamefont {T.~D.}\ \bibnamefont {Lee}}, \bibinfo {author} {\bibfnamefont {K.}~\bibnamefont {Huang}},\ and\ \bibinfo {author} {\bibfnamefont {C.~N.}\ \bibnamefont {Yang}},\ }\bibfield  {title} {\bibinfo {title} {Eigenvalues and eigenfunctions of a bose system of hard spheres and its low-temperature properties},\ }\href {https://doi.org/10.1103/PhysRev.106.1135} {\bibfield  {journal} {\bibinfo  {journal} {Phys. Rev.}\ }\textbf {\bibinfo {volume} {106}},\ \bibinfo {pages} {1135} (\bibinfo {year} {1957})}\BibitemShut {NoStop}%
\bibitem [{\citenamefont {Lima}\ and\ \citenamefont {Pelster}(2011)}]{Lima2011}%
  \BibitemOpen
  \bibfield  {author} {\bibinfo {author} {\bibfnamefont {A.~R.~P.}\ \bibnamefont {Lima}}\ and\ \bibinfo {author} {\bibfnamefont {A.}~\bibnamefont {Pelster}},\ }\bibfield  {title} {\bibinfo {title} {Quantum fluctuations in dipolar bose gases},\ }\href {https://doi.org/10.1103/PhysRevA.84.041604} {\bibfield  {journal} {\bibinfo  {journal} {Phys. Rev. A}\ }\textbf {\bibinfo {volume} {84}},\ \bibinfo {pages} {041604} (\bibinfo {year} {2011})}\BibitemShut {NoStop}%
\bibitem [{\citenamefont {Jain}\ \emph {et~al.}(2011)\citenamefont {Jain}, \citenamefont {Cinti},\ and\ \citenamefont {Boninsegni}}]{Jain2011}%
  \BibitemOpen
  \bibfield  {author} {\bibinfo {author} {\bibfnamefont {P.}~\bibnamefont {Jain}}, \bibinfo {author} {\bibfnamefont {F.}~\bibnamefont {Cinti}},\ and\ \bibinfo {author} {\bibfnamefont {M.}~\bibnamefont {Boninsegni}},\ }\bibfield  {title} {\bibinfo {title} {Structure, bose-einstein condensation, and superfluidity of two-dimensional confined dipolar assemblies},\ }\href {https://doi.org/10.1103/PhysRevB.84.014534} {\bibfield  {journal} {\bibinfo  {journal} {Phys. Rev. B}\ }\textbf {\bibinfo {volume} {84}},\ \bibinfo {pages} {014534} (\bibinfo {year} {2011})}\BibitemShut {NoStop}%
\bibitem [{\citenamefont {Cinti}\ and\ \citenamefont {Boninsegni}(2017)}]{Cinti2017a}%
  \BibitemOpen
  \bibfield  {author} {\bibinfo {author} {\bibfnamefont {F.}~\bibnamefont {Cinti}}\ and\ \bibinfo {author} {\bibfnamefont {M.}~\bibnamefont {Boninsegni}},\ }\bibfield  {title} {\bibinfo {title} {Classical and quantum filaments in the ground state of trapped dipolar bose gases},\ }\href {https://doi.org/10.1103/PhysRevA.96.013627} {\bibfield  {journal} {\bibinfo  {journal} {Phys. Rev. A}\ }\textbf {\bibinfo {volume} {96}},\ \bibinfo {pages} {013627} (\bibinfo {year} {2017})}\BibitemShut {NoStop}%
\bibitem [{\citenamefont {Cinti}\ \emph {et~al.}(2017{\natexlab{a}})\citenamefont {Cinti}, \citenamefont {Cappellaro}, \citenamefont {Salasnich},\ and\ \citenamefont {Macr\`{\i}}}]{Cinti2017b}%
  \BibitemOpen
  \bibfield  {author} {\bibinfo {author} {\bibfnamefont {F.}~\bibnamefont {Cinti}}, \bibinfo {author} {\bibfnamefont {A.}~\bibnamefont {Cappellaro}}, \bibinfo {author} {\bibfnamefont {L.}~\bibnamefont {Salasnich}},\ and\ \bibinfo {author} {\bibfnamefont {T.}~\bibnamefont {Macr\`{\i}}},\ }\bibfield  {title} {\bibinfo {title} {Superfluid filaments of dipolar bosons in free space},\ }\href {https://doi.org/10.1103/PhysRevLett.119.215302} {\bibfield  {journal} {\bibinfo  {journal} {Phys. Rev. Lett.}\ }\textbf {\bibinfo {volume} {119}},\ \bibinfo {pages} {215302} (\bibinfo {year} {2017}{\natexlab{a}})}\BibitemShut {NoStop}%
\bibitem [{\citenamefont {Saito}(2016)}]{Saito2016}%
  \BibitemOpen
  \bibfield  {author} {\bibinfo {author} {\bibfnamefont {H.}~\bibnamefont {Saito}},\ }\bibfield  {title} {\bibinfo {title} {Path-integral monte carlo study on a droplet of a dipolar bose–einstein condensate stabilized by quantum fluctuation},\ }\href {https://doi.org/10.7566/jpsj.85.053001} {\bibfield  {journal} {\bibinfo  {journal} {J. Phys. Soc. JPN}\ }\textbf {\bibinfo {volume} {85}},\ \bibinfo {pages} {053001} (\bibinfo {year} {2016})}\BibitemShut {NoStop}%
\bibitem [{\citenamefont {Macia}\ \emph {et~al.}(2016)\citenamefont {Macia}, \citenamefont {S\'anchez-Baena}, \citenamefont {Boronat},\ and\ \citenamefont {Mazzanti}}]{Macia2016}%
  \BibitemOpen
  \bibfield  {author} {\bibinfo {author} {\bibfnamefont {A.}~\bibnamefont {Macia}}, \bibinfo {author} {\bibfnamefont {J.}~\bibnamefont {S\'anchez-Baena}}, \bibinfo {author} {\bibfnamefont {J.}~\bibnamefont {Boronat}},\ and\ \bibinfo {author} {\bibfnamefont {F.}~\bibnamefont {Mazzanti}},\ }\bibfield  {title} {\bibinfo {title} {Droplets of trapped quantum dipolar bosons},\ }\href {https://doi.org/10.1103/PhysRevLett.117.205301} {\bibfield  {journal} {\bibinfo  {journal} {Phys. Rev. Lett.}\ }\textbf {\bibinfo {volume} {117}},\ \bibinfo {pages} {205301} (\bibinfo {year} {2016})}\BibitemShut {NoStop}%
\bibitem [{\citenamefont {B\"ottcher}\ \emph {et~al.}(2019{\natexlab{b}})\citenamefont {B\"ottcher}, \citenamefont {Wenzel}, \citenamefont {Schmidt}, \citenamefont {Guo}, \citenamefont {Langen}, \citenamefont {Ferrier-Barbut}, \citenamefont {Pfau}, \citenamefont {Bomb\'{\i}n}, \citenamefont {S\'anchez-Baena}, \citenamefont {Boronat},\ and\ \citenamefont {Mazzanti}}]{Bottcher2019b}%
  \BibitemOpen
  \bibfield  {author} {\bibinfo {author} {\bibfnamefont {F.}~\bibnamefont {B\"ottcher}}, \bibinfo {author} {\bibfnamefont {M.}~\bibnamefont {Wenzel}}, \bibinfo {author} {\bibfnamefont {J.}~\bibnamefont {Schmidt}}, \bibinfo {author} {\bibfnamefont {M.}~\bibnamefont {Guo}}, \bibinfo {author} {\bibfnamefont {T.}~\bibnamefont {Langen}}, \bibinfo {author} {\bibfnamefont {I.}~\bibnamefont {Ferrier-Barbut}}, \bibinfo {author} {\bibfnamefont {T.}~\bibnamefont {Pfau}}, \bibinfo {author} {\bibfnamefont {R.}~\bibnamefont {Bomb\'{\i}n}}, \bibinfo {author} {\bibfnamefont {J.}~\bibnamefont {S\'anchez-Baena}}, \bibinfo {author} {\bibfnamefont {J.}~\bibnamefont {Boronat}},\ and\ \bibinfo {author} {\bibfnamefont {F.}~\bibnamefont {Mazzanti}},\ }\bibfield  {title} {\bibinfo {title} {Dilute dipolar quantum droplets beyond the extended gross-pitaevskii equation},\ }\href {https://doi.org/10.1103/PhysRevResearch.1.033088} {\bibfield  {journal} {\bibinfo  {journal} {Phys. Rev. Res.}\ }\textbf {\bibinfo {volume} {1}},\ \bibinfo
  {pages} {033088} (\bibinfo {year} {2019}{\natexlab{b}})}\BibitemShut {NoStop}%
\bibitem [{\citenamefont {Boninsegni}(2021)}]{Boninsegni2021}%
  \BibitemOpen
  \bibfield  {author} {\bibinfo {author} {\bibfnamefont {M.}~\bibnamefont {Boninsegni}},\ }\bibfield  {title} {\bibinfo {title} {Morphology of dipolar bose droplets},\ }\href {https://doi.org/https://doi.org/10.1016/j.rinp.2021.104935} {\bibfield  {journal} {\bibinfo  {journal} {Results in Physics}\ }\textbf {\bibinfo {volume} {31}},\ \bibinfo {pages} {104935} (\bibinfo {year} {2021})}\BibitemShut {NoStop}%
\bibitem [{\citenamefont {Kora}\ and\ \citenamefont {Boninsegni}(2019)}]{Kora2019}%
  \BibitemOpen
  \bibfield  {author} {\bibinfo {author} {\bibfnamefont {Y.}~\bibnamefont {Kora}}\ and\ \bibinfo {author} {\bibfnamefont {M.}~\bibnamefont {Boninsegni}},\ }\bibfield  {title} {\bibinfo {title} {Patterned supersolids in dipolar bose systems},\ }\href {https://doi.org/10.1007/s10909-019-02229-z} {\bibfield  {journal} {\bibinfo  {journal} {J. Low Temp. Phys.}\ }\textbf {\bibinfo {volume} {197}},\ \bibinfo {pages} {337} (\bibinfo {year} {2019})}\BibitemShut {NoStop}%
\bibitem [{\citenamefont {Sohmen}\ \emph {et~al.}(2021)\citenamefont {Sohmen}, \citenamefont {Politi}, \citenamefont {Klaus}, \citenamefont {Chomaz}, \citenamefont {Mark}, \citenamefont {Norcia},\ and\ \citenamefont {Ferlaino}}]{Sohmen2021}%
  \BibitemOpen
  \bibfield  {author} {\bibinfo {author} {\bibfnamefont {M.}~\bibnamefont {Sohmen}}, \bibinfo {author} {\bibfnamefont {C.}~\bibnamefont {Politi}}, \bibinfo {author} {\bibfnamefont {L.}~\bibnamefont {Klaus}}, \bibinfo {author} {\bibfnamefont {L.}~\bibnamefont {Chomaz}}, \bibinfo {author} {\bibfnamefont {M.~J.}\ \bibnamefont {Mark}}, \bibinfo {author} {\bibfnamefont {M.~A.}\ \bibnamefont {Norcia}},\ and\ \bibinfo {author} {\bibfnamefont {F.}~\bibnamefont {Ferlaino}},\ }\bibfield  {title} {\bibinfo {title} {Birth, life, and death of a dipolar supersolid},\ }\href {https://doi.org/10.1103/PhysRevLett.126.233401} {\bibfield  {journal} {\bibinfo  {journal} {Phys. Rev. Lett.}\ }\textbf {\bibinfo {volume} {126}},\ \bibinfo {pages} {233401} (\bibinfo {year} {2021})}\BibitemShut {NoStop}%
\bibitem [{\citenamefont {Norcia}\ \emph {et~al.}(2021)\citenamefont {Norcia}, \citenamefont {Politi}, \citenamefont {Klaus}, \citenamefont {Poli}, \citenamefont {Sohmen}, \citenamefont {Mark}, \citenamefont {Bisset}, \citenamefont {Santos},\ and\ \citenamefont {Ferlaino}}]{Norcia2021}%
  \BibitemOpen
  \bibfield  {author} {\bibinfo {author} {\bibfnamefont {M.~A.}\ \bibnamefont {Norcia}}, \bibinfo {author} {\bibfnamefont {C.}~\bibnamefont {Politi}}, \bibinfo {author} {\bibfnamefont {L.}~\bibnamefont {Klaus}}, \bibinfo {author} {\bibfnamefont {E.}~\bibnamefont {Poli}}, \bibinfo {author} {\bibfnamefont {M.}~\bibnamefont {Sohmen}}, \bibinfo {author} {\bibfnamefont {M.~J.}\ \bibnamefont {Mark}}, \bibinfo {author} {\bibfnamefont {R.~N.}\ \bibnamefont {Bisset}}, \bibinfo {author} {\bibfnamefont {L.}~\bibnamefont {Santos}},\ and\ \bibinfo {author} {\bibfnamefont {F.}~\bibnamefont {Ferlaino}},\ }\bibfield  {title} {\bibinfo {title} {Two-dimensional supersolidity in a dipolar quantum gas},\ }\href {https://doi.org/10.1038/s41586-021-03725-7} {\bibfield  {journal} {\bibinfo  {journal} {Nature}\ }\textbf {\bibinfo {volume} {596}},\ \bibinfo {pages} {357} (\bibinfo {year} {2021})}\BibitemShut {NoStop}%
\bibitem [{\citenamefont {Sinha}\ and\ \citenamefont {Sinha}(2025)}]{Sinha2025}%
  \BibitemOpen
  \bibfield  {author} {\bibinfo {author} {\bibfnamefont {S.}~\bibnamefont {Sinha}}\ and\ \bibinfo {author} {\bibfnamefont {S.}~\bibnamefont {Sinha}},\ }\bibfield  {title} {\bibinfo {title} {Supersolid phases of bosons},\ }\bibfield  {journal} {\bibinfo  {journal} {Journal of Physics: Condensed Matter}\ }\href {https://doi.org/10.1088/1361-648x/adf6fb} {10.1088/1361-648x/adf6fb} (\bibinfo {year} {2025})\BibitemShut {NoStop}%
\bibitem [{\citenamefont {Recati}\ and\ \citenamefont {Stringari}(2023)}]{Recati2023}%
  \BibitemOpen
  \bibfield  {author} {\bibinfo {author} {\bibfnamefont {A.}~\bibnamefont {Recati}}\ and\ \bibinfo {author} {\bibfnamefont {S.}~\bibnamefont {Stringari}},\ }\bibfield  {title} {\bibinfo {title} {Supersolidity in ultracold dipolar gases},\ }\bibfield  {journal} {\bibinfo  {journal} {Nat. Rev. Phys.}\ }\href {https://doi.org/10.1038/s42254-023-00648-2} {10.1038/s42254-023-00648-2} (\bibinfo {year} {2023})\BibitemShut {NoStop}%
\bibitem [{\citenamefont {Chomaz}\ \emph {et~al.}(2022)\citenamefont {Chomaz}, \citenamefont {Ferrier-Barbut}, \citenamefont {Ferlaino}, \citenamefont {Laburthe-Tolra}, \citenamefont {Lev},\ and\ \citenamefont {Pfau}}]{Chomaz2022}%
  \BibitemOpen
  \bibfield  {author} {\bibinfo {author} {\bibfnamefont {L.}~\bibnamefont {Chomaz}}, \bibinfo {author} {\bibfnamefont {I.}~\bibnamefont {Ferrier-Barbut}}, \bibinfo {author} {\bibfnamefont {F.}~\bibnamefont {Ferlaino}}, \bibinfo {author} {\bibfnamefont {B.}~\bibnamefont {Laburthe-Tolra}}, \bibinfo {author} {\bibfnamefont {B.~L.}\ \bibnamefont {Lev}},\ and\ \bibinfo {author} {\bibfnamefont {T.}~\bibnamefont {Pfau}},\ }\bibfield  {title} {\bibinfo {title} {Dipolar physics: a review of experiments with magnetic quantum gases},\ }\href {https://doi.org/10.1088/1361-6633/aca814} {\bibfield  {journal} {\bibinfo  {journal} {Rep. Prog. Phys.}\ }\textbf {\bibinfo {volume} {86}},\ \bibinfo {pages} {026401} (\bibinfo {year} {2022})}\BibitemShut {NoStop}%
\bibitem [{\citenamefont {Schindewolf}\ \emph {et~al.}(2026)\citenamefont {Schindewolf}, \citenamefont {Hertkorn}, \citenamefont {Stevenson}, \citenamefont {Ciardi}, \citenamefont {Groß}, \citenamefont {Wang}, \citenamefont {Karman}, \citenamefont {Quéméner}, \citenamefont {Will}, \citenamefont {Pohl},\ and\ \citenamefont {Langen}}]{Schindewolf2026}%
  \BibitemOpen
  \bibfield  {author} {\bibinfo {author} {\bibfnamefont {A.}~\bibnamefont {Schindewolf}}, \bibinfo {author} {\bibfnamefont {J.}~\bibnamefont {Hertkorn}}, \bibinfo {author} {\bibfnamefont {I.}~\bibnamefont {Stevenson}}, \bibinfo {author} {\bibfnamefont {M.}~\bibnamefont {Ciardi}}, \bibinfo {author} {\bibfnamefont {P.}~\bibnamefont {Groß}}, \bibinfo {author} {\bibfnamefont {D.}~\bibnamefont {Wang}}, \bibinfo {author} {\bibfnamefont {T.}~\bibnamefont {Karman}}, \bibinfo {author} {\bibfnamefont {G.}~\bibnamefont {Quéméner}}, \bibinfo {author} {\bibfnamefont {S.}~\bibnamefont {Will}}, \bibinfo {author} {\bibfnamefont {T.}~\bibnamefont {Pohl}},\ and\ \bibinfo {author} {\bibfnamefont {T.}~\bibnamefont {Langen}},\ }\bibfield  {title} {\bibinfo {title} {Colloquium: Strongly dipolar molecular bose-einstein condensates: From few- to many-body physics},\ }\href {https://doi.org/10.1103/r55l-f93m} {\bibfield  {journal} {\bibinfo  {journal} {Rev. Mod. Phys.}\ } (\bibinfo {year} {2026})}\BibitemShut {NoStop}%
\bibitem [{\citenamefont {Karman}\ and\ \citenamefont {Hutson}(2018)}]{Karman2018}%
  \BibitemOpen
  \bibfield  {author} {\bibinfo {author} {\bibfnamefont {T.}~\bibnamefont {Karman}}\ and\ \bibinfo {author} {\bibfnamefont {J.~M.}\ \bibnamefont {Hutson}},\ }\bibfield  {title} {\bibinfo {title} {Microwave shielding of ultracold polar molecules},\ }\href {https://doi.org/10.1103/PhysRevLett.121.163401} {\bibfield  {journal} {\bibinfo  {journal} {Phys. Rev. Lett.}\ }\textbf {\bibinfo {volume} {121}},\ \bibinfo {pages} {163401} (\bibinfo {year} {2018})}\BibitemShut {NoStop}%
\bibitem [{\citenamefont {Andreev}\ and\ \citenamefont {Lifshitz}(1971)}]{Andreev1971}%
  \BibitemOpen
  \bibfield  {author} {\bibinfo {author} {\bibfnamefont {A.~F.}\ \bibnamefont {Andreev}}\ and\ \bibinfo {author} {\bibfnamefont {I.~M.}\ \bibnamefont {Lifshitz}},\ }\bibfield  {title} {\bibinfo {title} {Quantum theory of defects in crystals},\ }\href {https://doi.org/10.1070/pu1971v013n05abeh004235} {\bibfield  {journal} {\bibinfo  {journal} {Soviet Physics Uspekhi}\ }\textbf {\bibinfo {volume} {13}},\ \bibinfo {pages} {670–670} (\bibinfo {year} {1971})}\BibitemShut {NoStop}%
\bibitem [{\citenamefont {Cinti}\ \emph {et~al.}(2014)\citenamefont {Cinti}, \citenamefont {Macr{\`i}}, \citenamefont {Lechner}, \citenamefont {Pupillo},\ and\ \citenamefont {Pohl}}]{Cinti2014}%
  \BibitemOpen
  \bibfield  {author} {\bibinfo {author} {\bibfnamefont {F.}~\bibnamefont {Cinti}}, \bibinfo {author} {\bibfnamefont {T.}~\bibnamefont {Macr{\`i}}}, \bibinfo {author} {\bibfnamefont {W.}~\bibnamefont {Lechner}}, \bibinfo {author} {\bibfnamefont {G.}~\bibnamefont {Pupillo}},\ and\ \bibinfo {author} {\bibfnamefont {T.}~\bibnamefont {Pohl}},\ }\bibfield  {title} {\bibinfo {title} {Defect-induced supersolidity with soft-core bosons},\ }\href {https://doi.org/10.1038/ncomms4235} {\bibfield  {journal} {\bibinfo  {journal} {Nat. Commun.}\ }\textbf {\bibinfo {volume} {5}},\ \bibinfo {pages} {3235} (\bibinfo {year} {2014})}\BibitemShut {NoStop}%
\bibitem [{\citenamefont {Bigagli}\ \emph {et~al.}(2024)\citenamefont {Bigagli}, \citenamefont {Yuan}, \citenamefont {Zhang}, \citenamefont {Bulatovic}, \citenamefont {Karman}, \citenamefont {Stevenson},\ and\ \citenamefont {Will}}]{Bigagli2024}%
  \BibitemOpen
  \bibfield  {author} {\bibinfo {author} {\bibfnamefont {N.}~\bibnamefont {Bigagli}}, \bibinfo {author} {\bibfnamefont {W.}~\bibnamefont {Yuan}}, \bibinfo {author} {\bibfnamefont {S.}~\bibnamefont {Zhang}}, \bibinfo {author} {\bibfnamefont {B.}~\bibnamefont {Bulatovic}}, \bibinfo {author} {\bibfnamefont {T.}~\bibnamefont {Karman}}, \bibinfo {author} {\bibfnamefont {I.}~\bibnamefont {Stevenson}},\ and\ \bibinfo {author} {\bibfnamefont {S.}~\bibnamefont {Will}},\ }\bibfield  {title} {\bibinfo {title} {Observation of bose–einstein condensation of dipolar molecules},\ }\href {https://doi.org/10.1038/s41586-024-07492-z} {\bibfield  {journal} {\bibinfo  {journal} {Nature}\ }\textbf {\bibinfo {volume} {631}},\ \bibinfo {pages} {289–293} (\bibinfo {year} {2024})}\BibitemShut {NoStop}%
\bibitem [{\citenamefont {Shi}\ \emph {et~al.}(2025)\citenamefont {Shi}, \citenamefont {Huang}, \citenamefont {Deng}, \citenamefont {Jin}, \citenamefont {Yi}, \citenamefont {Shi},\ and\ \citenamefont {Wang}}]{Shi2025}%
  \BibitemOpen
  \bibfield  {author} {\bibinfo {author} {\bibfnamefont {Z.}~\bibnamefont {Shi}}, \bibinfo {author} {\bibfnamefont {Z.}~\bibnamefont {Huang}}, \bibinfo {author} {\bibfnamefont {F.}~\bibnamefont {Deng}}, \bibinfo {author} {\bibfnamefont {W.-J.}\ \bibnamefont {Jin}}, \bibinfo {author} {\bibfnamefont {S.}~\bibnamefont {Yi}}, \bibinfo {author} {\bibfnamefont {T.}~\bibnamefont {Shi}},\ and\ \bibinfo {author} {\bibfnamefont {D.}~\bibnamefont {Wang}},\ }\href {https://arxiv.org/abs/2508.20518} {\bibinfo {title} {Bose-einstein condensate of ultracold sodium-rubidium molecules with tunable dipolar interactions}} (\bibinfo {year} {2025}),\ \Eprint {https://arxiv.org/abs/2508.20518} {arXiv:2508.20518 [cond-mat.quant-gas]} \BibitemShut {NoStop}%
\bibitem [{\citenamefont {S\'anchez-Baena}\ \emph {et~al.}(2025)\citenamefont {S\'anchez-Baena}, \citenamefont {Pascual}, \citenamefont {Bomb\'{\i}n}, \citenamefont {Mazzanti},\ and\ \citenamefont {Boronat}}]{Baena2025}%
  \BibitemOpen
  \bibfield  {author} {\bibinfo {author} {\bibfnamefont {J.}~\bibnamefont {S\'anchez-Baena}}, \bibinfo {author} {\bibfnamefont {G.}~\bibnamefont {Pascual}}, \bibinfo {author} {\bibfnamefont {R.}~\bibnamefont {Bomb\'{\i}n}}, \bibinfo {author} {\bibfnamefont {F.}~\bibnamefont {Mazzanti}},\ and\ \bibinfo {author} {\bibfnamefont {J.}~\bibnamefont {Boronat}},\ }\bibfield  {title} {\bibinfo {title} {Thermal behavior of bose-einstein condensates of polar molecules},\ }\href {https://doi.org/10.1103/9sbg-6qqw} {\bibfield  {journal} {\bibinfo  {journal} {Phys. Rev. Res.}\ }\textbf {\bibinfo {volume} {7}},\ \bibinfo {pages} {033080} (\bibinfo {year} {2025})}\BibitemShut {NoStop}%
\bibitem [{\citenamefont {Dutta}\ \emph {et~al.}(2025)\citenamefont {Dutta}, \citenamefont {Mukherjee},\ and\ \citenamefont {Hutson}}]{Dutta2025}%
  \BibitemOpen
  \bibfield  {author} {\bibinfo {author} {\bibfnamefont {J.}~\bibnamefont {Dutta}}, \bibinfo {author} {\bibfnamefont {B.}~\bibnamefont {Mukherjee}},\ and\ \bibinfo {author} {\bibfnamefont {J.~M.}\ \bibnamefont {Hutson}},\ }\bibfield  {title} {\bibinfo {title} {Universality in the microwave shielding of ultracold polar molecules},\ }\href {https://doi.org/10.1103/PhysRevResearch.7.023164} {\bibfield  {journal} {\bibinfo  {journal} {Phys. Rev. Res.}\ }\textbf {\bibinfo {volume} {7}},\ \bibinfo {pages} {023164} (\bibinfo {year} {2025})}\BibitemShut {NoStop}%
\bibitem [{\citenamefont {B\"uchler}\ \emph {et~al.}(2007)\citenamefont {B\"uchler}, \citenamefont {Demler}, \citenamefont {Lukin}, \citenamefont {Micheli}, \citenamefont {Prokof'ev}, \citenamefont {Pupillo},\ and\ \citenamefont {Zoller}}]{Buchler2007}%
  \BibitemOpen
  \bibfield  {author} {\bibinfo {author} {\bibfnamefont {H.~P.}\ \bibnamefont {B\"uchler}}, \bibinfo {author} {\bibfnamefont {E.}~\bibnamefont {Demler}}, \bibinfo {author} {\bibfnamefont {M.}~\bibnamefont {Lukin}}, \bibinfo {author} {\bibfnamefont {A.}~\bibnamefont {Micheli}}, \bibinfo {author} {\bibfnamefont {N.}~\bibnamefont {Prokof'ev}}, \bibinfo {author} {\bibfnamefont {G.}~\bibnamefont {Pupillo}},\ and\ \bibinfo {author} {\bibfnamefont {P.}~\bibnamefont {Zoller}},\ }\bibfield  {title} {\bibinfo {title} {Strongly correlated 2d quantum phases with cold polar molecules: Controlling the shape of the interaction potential},\ }\href {https://doi.org/10.1103/PhysRevLett.98.060404} {\bibfield  {journal} {\bibinfo  {journal} {Phys. Rev. Lett.}\ }\textbf {\bibinfo {volume} {98}},\ \bibinfo {pages} {060404} (\bibinfo {year} {2007})}\BibitemShut {NoStop}%
\bibitem [{\citenamefont {Micheli}\ \emph {et~al.}(2007)\citenamefont {Micheli}, \citenamefont {Pupillo}, \citenamefont {B\"uchler},\ and\ \citenamefont {Zoller}}]{Michelli2007}%
  \BibitemOpen
  \bibfield  {author} {\bibinfo {author} {\bibfnamefont {A.}~\bibnamefont {Micheli}}, \bibinfo {author} {\bibfnamefont {G.}~\bibnamefont {Pupillo}}, \bibinfo {author} {\bibfnamefont {H.~P.}\ \bibnamefont {B\"uchler}},\ and\ \bibinfo {author} {\bibfnamefont {P.}~\bibnamefont {Zoller}},\ }\bibfield  {title} {\bibinfo {title} {Cold polar molecules in two-dimensional traps: Tailoring interactions with external fields for novel quantum phases},\ }\href {https://doi.org/10.1103/PhysRevA.76.043604} {\bibfield  {journal} {\bibinfo  {journal} {Phys. Rev. A}\ }\textbf {\bibinfo {volume} {76}},\ \bibinfo {pages} {043604} (\bibinfo {year} {2007})}\BibitemShut {NoStop}%
\bibitem [{\citenamefont {Lassabli\`ere}\ and\ \citenamefont {Qu\'em\'ener}(2018)}]{Lassabliere2018}%
  \BibitemOpen
  \bibfield  {author} {\bibinfo {author} {\bibfnamefont {L.}~\bibnamefont {Lassabli\`ere}}\ and\ \bibinfo {author} {\bibfnamefont {G.}~\bibnamefont {Qu\'em\'ener}},\ }\bibfield  {title} {\bibinfo {title} {Controlling the scattering length of ultracold dipolar molecules},\ }\href {https://doi.org/10.1103/PhysRevLett.121.163402} {\bibfield  {journal} {\bibinfo  {journal} {Phys. Rev. Lett.}\ }\textbf {\bibinfo {volume} {121}},\ \bibinfo {pages} {163402} (\bibinfo {year} {2018})}\BibitemShut {NoStop}%
\bibitem [{\citenamefont {Deng}\ \emph {et~al.}(2025)\citenamefont {Deng}, \citenamefont {Hu}, \citenamefont {Jin}, \citenamefont {Yi},\ and\ \citenamefont {Shi}}]{Deng2025}%
  \BibitemOpen
  \bibfield  {author} {\bibinfo {author} {\bibfnamefont {F.}~\bibnamefont {Deng}}, \bibinfo {author} {\bibfnamefont {X.}~\bibnamefont {Hu}}, \bibinfo {author} {\bibfnamefont {W.-J.}\ \bibnamefont {Jin}}, \bibinfo {author} {\bibfnamefont {S.}~\bibnamefont {Yi}},\ and\ \bibinfo {author} {\bibfnamefont {T.}~\bibnamefont {Shi}},\ }\bibfield  {title} {\bibinfo {title} {Two- and many-body physics of ultracold molecules dressed by dual microwave fields},\ }\bibfield  {journal} {\bibinfo  {journal} {Nature Communications}\ }\textbf {\bibinfo {volume} {16}},\ \href {https://doi.org/10.1038/s41467-025-66067-2} {10.1038/s41467-025-66067-2} (\bibinfo {year} {2025})\BibitemShut {NoStop}%
\bibitem [{\citenamefont {Langen}\ \emph {et~al.}(2025)\citenamefont {Langen}, \citenamefont {Boronat}, \citenamefont {S\'anchez-Baena}, \citenamefont {Bomb\'{\i}n}, \citenamefont {Karman},\ and\ \citenamefont {Mazzanti}}]{Langen2025}%
  \BibitemOpen
  \bibfield  {author} {\bibinfo {author} {\bibfnamefont {T.}~\bibnamefont {Langen}}, \bibinfo {author} {\bibfnamefont {J.}~\bibnamefont {Boronat}}, \bibinfo {author} {\bibfnamefont {J.}~\bibnamefont {S\'anchez-Baena}}, \bibinfo {author} {\bibfnamefont {R.}~\bibnamefont {Bomb\'{\i}n}}, \bibinfo {author} {\bibfnamefont {T.}~\bibnamefont {Karman}},\ and\ \bibinfo {author} {\bibfnamefont {F.}~\bibnamefont {Mazzanti}},\ }\bibfield  {title} {\bibinfo {title} {Dipolar droplets of strongly interacting molecules},\ }\href {https://doi.org/10.1103/PhysRevLett.134.053001} {\bibfield  {journal} {\bibinfo  {journal} {Phys. Rev. Lett.}\ }\textbf {\bibinfo {volume} {134}},\ \bibinfo {pages} {053001} (\bibinfo {year} {2025})}\BibitemShut {NoStop}%
\bibitem [{\citenamefont {Ciardi}\ \emph {et~al.}(2025{\natexlab{a}})\citenamefont {Ciardi}, \citenamefont {Pedersen}, \citenamefont {Langen},\ and\ \citenamefont {Pohl}}]{Ciardi2025}%
  \BibitemOpen
  \bibfield  {author} {\bibinfo {author} {\bibfnamefont {M.}~\bibnamefont {Ciardi}}, \bibinfo {author} {\bibfnamefont {K.~R.}\ \bibnamefont {Pedersen}}, \bibinfo {author} {\bibfnamefont {T.}~\bibnamefont {Langen}},\ and\ \bibinfo {author} {\bibfnamefont {T.}~\bibnamefont {Pohl}},\ }\bibfield  {title} {\bibinfo {title} {Self-bound superfluid membranes and monolayer crystals of ultracold polar molecules},\ }\href {https://doi.org/10.1103/v7gw-xy36} {\bibfield  {journal} {\bibinfo  {journal} {Phys. Rev. Lett.}\ }\textbf {\bibinfo {volume} {135}},\ \bibinfo {pages} {153401} (\bibinfo {year} {2025}{\natexlab{a}})}\BibitemShut {NoStop}%
\bibitem [{\citenamefont {Wang}(2007)}]{wang07}%
  \BibitemOpen
  \bibfield  {author} {\bibinfo {author} {\bibfnamefont {D.-W.}\ \bibnamefont {Wang}},\ }\bibfield  {title} {\bibinfo {title} {Quantum phase transitions of polar molecules in bilayer systems},\ }\href {https://doi.org/10.1103/PhysRevLett.98.060403} {\bibfield  {journal} {\bibinfo  {journal} {Phys. Rev. Lett.}\ }\textbf {\bibinfo {volume} {98}},\ \bibinfo {pages} {060403} (\bibinfo {year} {2007})}\BibitemShut {NoStop}%
\bibitem [{\citenamefont {Rydow}\ \emph {et~al.}(2025)\citenamefont {Rydow}, \citenamefont {Singh}, \citenamefont {Beregi}, \citenamefont {Chang}, \citenamefont {Mathey}, \citenamefont {Foot},\ and\ \citenamefont {Sunami}}]{Rydow2025}%
  \BibitemOpen
  \bibfield  {author} {\bibinfo {author} {\bibfnamefont {E.}~\bibnamefont {Rydow}}, \bibinfo {author} {\bibfnamefont {V.~P.}\ \bibnamefont {Singh}}, \bibinfo {author} {\bibfnamefont {A.}~\bibnamefont {Beregi}}, \bibinfo {author} {\bibfnamefont {E.}~\bibnamefont {Chang}}, \bibinfo {author} {\bibfnamefont {L.}~\bibnamefont {Mathey}}, \bibinfo {author} {\bibfnamefont {C.~J.}\ \bibnamefont {Foot}},\ and\ \bibinfo {author} {\bibfnamefont {S.}~\bibnamefont {Sunami}},\ }\bibfield  {title} {\bibinfo {title} {Observation of a bilayer superfluid with interlayer coherence},\ }\bibfield  {journal} {\bibinfo  {journal} {Nature Communications}\ }\textbf {\bibinfo {volume} {16}},\ \href {https://doi.org/10.1038/s41467-025-62277-w} {10.1038/s41467-025-62277-w} (\bibinfo {year} {2025})\BibitemShut {NoStop}%
\bibitem [{\citenamefont {Cinti}\ \emph {et~al.}(2026)\citenamefont {Cinti}, \citenamefont {Ciardi}, \citenamefont {Prestipino},\ and\ \citenamefont {Pellicane}}]{Cinti2026}%
  \BibitemOpen
  \bibfield  {author} {\bibinfo {author} {\bibfnamefont {F.}~\bibnamefont {Cinti}}, \bibinfo {author} {\bibfnamefont {M.}~\bibnamefont {Ciardi}}, \bibinfo {author} {\bibfnamefont {S.}~\bibnamefont {Prestipino}},\ and\ \bibinfo {author} {\bibfnamefont {G.}~\bibnamefont {Pellicane}},\ }\href {https://arxiv.org/abs/2603.04650} {\bibinfo {title} {Layering and superfluidity of soft-core bosons in shallow spherical traps}} (\bibinfo {year} {2026}),\ \Eprint {https://arxiv.org/abs/2603.04650} {arXiv:2603.04650 [cond-mat.quant-gas]} \BibitemShut {NoStop}%
\bibitem [{\citenamefont {Eisenstein}(1996)}]{Eisenstein1996}%
  \BibitemOpen
  \bibfield  {author} {\bibinfo {author} {\bibfnamefont {J.~P.}\ \bibnamefont {Eisenstein}},\ }\bibinfo {title} {Experimental studies of multicomponent quantum hall systems},\ in\ \href {https://doi.org/https://doi.org/10.1002/9783527617258.ch2} {\emph {\bibinfo {booktitle} {Perspectives in Quantum Hall Effects}}}\ (\bibinfo  {publisher} {John Wiley \& Sons, Ltd},\ \bibinfo {year} {1996})\ Chap.~\bibinfo {chapter} {2}, pp.\ \bibinfo {pages} {37--70}\BibitemShut {NoStop}%
\bibitem [{\citenamefont {Girvin}\ and\ \citenamefont {MacDonald}(1996)}]{Girvin1996}%
  \BibitemOpen
  \bibfield  {author} {\bibinfo {author} {\bibfnamefont {S.~M.}\ \bibnamefont {Girvin}}\ and\ \bibinfo {author} {\bibfnamefont {A.~H.}\ \bibnamefont {MacDonald}},\ }\bibinfo {title} {Multicomponent quantum hall systems: The sum of their parts and more},\ in\ \href {https://doi.org/https://doi.org/10.1002/9783527617258.ch5} {\emph {\bibinfo {booktitle} {Perspectives in Quantum Hall Effects}}}\ (\bibinfo  {publisher} {John Wiley \& Sons, Ltd},\ \bibinfo {year} {1996})\ Chap.~\bibinfo {chapter} {5}, pp.\ \bibinfo {pages} {161--224}\BibitemShut {NoStop}%
\bibitem [{\citenamefont {Perali}\ \emph {et~al.}(2013)\citenamefont {Perali}, \citenamefont {Neilson},\ and\ \citenamefont {Hamilton}}]{Perali2013}%
  \BibitemOpen
  \bibfield  {author} {\bibinfo {author} {\bibfnamefont {A.}~\bibnamefont {Perali}}, \bibinfo {author} {\bibfnamefont {D.}~\bibnamefont {Neilson}},\ and\ \bibinfo {author} {\bibfnamefont {A.~R.}\ \bibnamefont {Hamilton}},\ }\bibfield  {title} {\bibinfo {title} {High-temperature superfluidity in double-bilayer graphene},\ }\href {https://doi.org/10.1103/PhysRevLett.110.146803} {\bibfield  {journal} {\bibinfo  {journal} {Phys. Rev. Lett.}\ }\textbf {\bibinfo {volume} {110}},\ \bibinfo {pages} {146803} (\bibinfo {year} {2013})}\BibitemShut {NoStop}%
\bibitem [{\citenamefont {Gao}\ \emph {et~al.}(2023)\citenamefont {Gao}, \citenamefont {Chan}, \citenamefont {Wang}, \citenamefont {Zhang}, \citenamefont {Jinxu}, \citenamefont {Cui}, \citenamefont {Yang}, \citenamefont {Liu}, \citenamefont {Shen}, \citenamefont {Sun}, \citenamefont {Jiang}, \citenamefont {Chiang},\ and\ \citenamefont {Chen}}]{Gao2023}%
  \BibitemOpen
  \bibfield  {author} {\bibinfo {author} {\bibfnamefont {Q.}~\bibnamefont {Gao}}, \bibinfo {author} {\bibfnamefont {Y.-h.}\ \bibnamefont {Chan}}, \bibinfo {author} {\bibfnamefont {Y.}~\bibnamefont {Wang}}, \bibinfo {author} {\bibfnamefont {H.}~\bibnamefont {Zhang}}, \bibinfo {author} {\bibfnamefont {P.}~\bibnamefont {Jinxu}}, \bibinfo {author} {\bibfnamefont {S.}~\bibnamefont {Cui}}, \bibinfo {author} {\bibfnamefont {Y.}~\bibnamefont {Yang}}, \bibinfo {author} {\bibfnamefont {Z.}~\bibnamefont {Liu}}, \bibinfo {author} {\bibfnamefont {D.}~\bibnamefont {Shen}}, \bibinfo {author} {\bibfnamefont {Z.}~\bibnamefont {Sun}}, \bibinfo {author} {\bibfnamefont {J.}~\bibnamefont {Jiang}}, \bibinfo {author} {\bibfnamefont {T.~C.}\ \bibnamefont {Chiang}},\ and\ \bibinfo {author} {\bibfnamefont {P.}~\bibnamefont {Chen}},\ }\bibfield  {title} {\bibinfo {title} {Evidence of high-temperature exciton condensation in a two-dimensional semimetal},\ }\bibfield  {journal} {\bibinfo  {journal} {Nature Communications}\ }\textbf
  {\bibinfo {volume} {14}},\ \href {https://doi.org/10.1038/s41467-023-36667-x} {10.1038/s41467-023-36667-x} (\bibinfo {year} {2023})\BibitemShut {NoStop}%
\bibitem [{\citenamefont {Deng}\ \emph {et~al.}(2023)\citenamefont {Deng}, \citenamefont {Chen}, \citenamefont {Luo}, \citenamefont {Zhang}, \citenamefont {Yi},\ and\ \citenamefont {Shi}}]{Deng2023}%
  \BibitemOpen
  \bibfield  {author} {\bibinfo {author} {\bibfnamefont {F.}~\bibnamefont {Deng}}, \bibinfo {author} {\bibfnamefont {X.-Y.}\ \bibnamefont {Chen}}, \bibinfo {author} {\bibfnamefont {X.-Y.}\ \bibnamefont {Luo}}, \bibinfo {author} {\bibfnamefont {W.}~\bibnamefont {Zhang}}, \bibinfo {author} {\bibfnamefont {S.}~\bibnamefont {Yi}},\ and\ \bibinfo {author} {\bibfnamefont {T.}~\bibnamefont {Shi}},\ }\bibfield  {title} {\bibinfo {title} {Effective potential and superfluidity of microwave-shielded polar molecules},\ }\href {https://doi.org/10.1103/PhysRevLett.130.183001} {\bibfield  {journal} {\bibinfo  {journal} {Phys. Rev. Lett.}\ }\textbf {\bibinfo {volume} {130}},\ \bibinfo {pages} {183001} (\bibinfo {year} {2023})}\BibitemShut {NoStop}%
\bibitem [{\citenamefont {Ceperley}(1995)}]{Ceperley1995}%
  \BibitemOpen
  \bibfield  {author} {\bibinfo {author} {\bibfnamefont {D.~M.}\ \bibnamefont {Ceperley}},\ }\bibfield  {title} {\bibinfo {title} {Path integrals in the theory of condensed helium},\ }\href {https://doi.org/10.1103/RevModPhys.67.279} {\bibfield  {journal} {\bibinfo  {journal} {Rev. Mod. Phys.}\ }\textbf {\bibinfo {volume} {67}},\ \bibinfo {pages} {279} (\bibinfo {year} {1995})}\BibitemShut {NoStop}%
\bibitem [{\citenamefont {Boninsegni}\ \emph {et~al.}(2006)\citenamefont {Boninsegni}, \citenamefont {Prokof'ev},\ and\ \citenamefont {Svistunov}}]{Boninsegni2006}%
  \BibitemOpen
  \bibfield  {author} {\bibinfo {author} {\bibfnamefont {M.}~\bibnamefont {Boninsegni}}, \bibinfo {author} {\bibfnamefont {N.}~\bibnamefont {Prokof'ev}},\ and\ \bibinfo {author} {\bibfnamefont {B.}~\bibnamefont {Svistunov}},\ }\bibfield  {title} {\bibinfo {title} {Worm algorithm for continuous-space path integral monte carlo simulations},\ }\href {https://doi.org/10.1103/PhysRevLett.96.070601} {\bibfield  {journal} {\bibinfo  {journal} {Phys. Rev. Lett.}\ }\textbf {\bibinfo {volume} {96}},\ \bibinfo {pages} {070601} (\bibinfo {year} {2006})}\BibitemShut {NoStop}%
\bibitem [{\citenamefont {Pollock}\ and\ \citenamefont {Ceperley}(1987)}]{Pollock1987}%
  \BibitemOpen
  \bibfield  {author} {\bibinfo {author} {\bibfnamefont {E.~L.}\ \bibnamefont {Pollock}}\ and\ \bibinfo {author} {\bibfnamefont {D.~M.}\ \bibnamefont {Ceperley}},\ }\bibfield  {title} {\bibinfo {title} {Path-integral computation of superfluid densities},\ }\href {https://doi.org/10.1103/PhysRevB.36.8343} {\bibfield  {journal} {\bibinfo  {journal} {Phys. Rev. B}\ }\textbf {\bibinfo {volume} {36}},\ \bibinfo {pages} {8343} (\bibinfo {year} {1987})}\BibitemShut {NoStop}%
\bibitem [{\citenamefont {Cinti}\ \emph {et~al.}(2017{\natexlab{b}})\citenamefont {Cinti}, \citenamefont {Cappellaro}, \citenamefont {Salasnich},\ and\ \citenamefont {Macr\`{\i}}}]{PhysRevLett.119.215302}%
  \BibitemOpen
  \bibfield  {author} {\bibinfo {author} {\bibfnamefont {F.}~\bibnamefont {Cinti}}, \bibinfo {author} {\bibfnamefont {A.}~\bibnamefont {Cappellaro}}, \bibinfo {author} {\bibfnamefont {L.}~\bibnamefont {Salasnich}},\ and\ \bibinfo {author} {\bibfnamefont {T.}~\bibnamefont {Macr\`{\i}}},\ }\bibfield  {title} {\bibinfo {title} {Superfluid filaments of dipolar bosons in free space},\ }\href {https://doi.org/10.1103/PhysRevLett.119.215302} {\bibfield  {journal} {\bibinfo  {journal} {Phys. Rev. Lett.}\ }\textbf {\bibinfo {volume} {119}},\ \bibinfo {pages} {215302} (\bibinfo {year} {2017}{\natexlab{b}})}\BibitemShut {NoStop}%
\bibitem [{\citenamefont {Ciardi}\ \emph {et~al.}(2025{\natexlab{b}})\citenamefont {Ciardi}, \citenamefont {Cinti}, \citenamefont {Pellicane},\ and\ \citenamefont {Prestipino}}]{ciardi2025b}%
  \BibitemOpen
  \bibfield  {author} {\bibinfo {author} {\bibfnamefont {M.}~\bibnamefont {Ciardi}}, \bibinfo {author} {\bibfnamefont {F.}~\bibnamefont {Cinti}}, \bibinfo {author} {\bibfnamefont {G.}~\bibnamefont {Pellicane}},\ and\ \bibinfo {author} {\bibfnamefont {S.}~\bibnamefont {Prestipino}},\ }\bibfield  {title} {\bibinfo {title} {Effects of gravity on supersolid order in bubble-trapped bosons},\ }\href {https://doi.org/10.1103/PhysRevB.111.024512} {\bibfield  {journal} {\bibinfo  {journal} {Phys. Rev. B.}\ }\textbf {\bibinfo {volume} {111}},\ \bibinfo {pages} {024512} (\bibinfo {year} {2025}{\natexlab{b}})}\BibitemShut {NoStop}%
\bibitem [{\citenamefont {Lu}\ \emph {et~al.}(2008)\citenamefont {Lu}, \citenamefont {Wu}, \citenamefont {Micheli},\ and\ \citenamefont {Pupillo}}]{Lu2008}%
  \BibitemOpen
  \bibfield  {author} {\bibinfo {author} {\bibfnamefont {X.}~\bibnamefont {Lu}}, \bibinfo {author} {\bibfnamefont {C.-Q.}\ \bibnamefont {Wu}}, \bibinfo {author} {\bibfnamefont {A.}~\bibnamefont {Micheli}},\ and\ \bibinfo {author} {\bibfnamefont {G.}~\bibnamefont {Pupillo}},\ }\bibfield  {title} {\bibinfo {title} {Structure and melting behavior of classical bilayer crystals of dipoles},\ }\href {https://doi.org/10.1103/PhysRevB.78.024108} {\bibfield  {journal} {\bibinfo  {journal} {Phys. Rev. B}\ }\textbf {\bibinfo {volume} {78}},\ \bibinfo {pages} {024108} (\bibinfo {year} {2008})}\BibitemShut {NoStop}%
\bibitem [{\citenamefont {Cinti}\ \emph {et~al.}(2017{\natexlab{c}})\citenamefont {Cinti}, \citenamefont {Wang},\ and\ \citenamefont {Boninsegni}}]{Cinti2017}%
  \BibitemOpen
  \bibfield  {author} {\bibinfo {author} {\bibfnamefont {F.}~\bibnamefont {Cinti}}, \bibinfo {author} {\bibfnamefont {D.-W.}\ \bibnamefont {Wang}},\ and\ \bibinfo {author} {\bibfnamefont {M.}~\bibnamefont {Boninsegni}},\ }\bibfield  {title} {\bibinfo {title} {Phases of dipolar bosons in a bilayer geometry},\ }\href {https://doi.org/10.1103/PhysRevA.95.023622} {\bibfield  {journal} {\bibinfo  {journal} {Phys. Rev. A}\ }\textbf {\bibinfo {volume} {95}},\ \bibinfo {pages} {023622} (\bibinfo {year} {2017}{\natexlab{c}})}\BibitemShut {NoStop}%
\bibitem [{\citenamefont {Sánchez-Baena}\ \emph {et~al.}(2023)\citenamefont {Sánchez-Baena}, \citenamefont {Politi}, \citenamefont {Maucher}, \citenamefont {Ferlaino},\ and\ \citenamefont {Pohl}}]{SnchezBaena2023}%
  \BibitemOpen
  \bibfield  {author} {\bibinfo {author} {\bibfnamefont {J.}~\bibnamefont {Sánchez-Baena}}, \bibinfo {author} {\bibfnamefont {C.}~\bibnamefont {Politi}}, \bibinfo {author} {\bibfnamefont {F.}~\bibnamefont {Maucher}}, \bibinfo {author} {\bibfnamefont {F.}~\bibnamefont {Ferlaino}},\ and\ \bibinfo {author} {\bibfnamefont {T.}~\bibnamefont {Pohl}},\ }\bibfield  {title} {\bibinfo {title} {Heating a dipolar quantum fluid into a solid},\ }\bibfield  {journal} {\bibinfo  {journal} {Nature Communications}\ }\textbf {\bibinfo {volume} {14}},\ \href {https://doi.org/10.1038/s41467-023-37207-3} {10.1038/s41467-023-37207-3} (\bibinfo {year} {2023})\BibitemShut {NoStop}%
\bibitem [{\citenamefont {Zampronio}\ \emph {et~al.}(2024)\citenamefont {Zampronio}, \citenamefont {Mendoza-Coto}, \citenamefont {Macr\`{\i}},\ and\ \citenamefont {Cinti}}]{Zampronio2024}%
  \BibitemOpen
  \bibfield  {author} {\bibinfo {author} {\bibfnamefont {V.}~\bibnamefont {Zampronio}}, \bibinfo {author} {\bibfnamefont {A.}~\bibnamefont {Mendoza-Coto}}, \bibinfo {author} {\bibfnamefont {T.}~\bibnamefont {Macr\`{\i}}},\ and\ \bibinfo {author} {\bibfnamefont {F.}~\bibnamefont {Cinti}},\ }\bibfield  {title} {\bibinfo {title} {Exploring quantum phases of dipolar gases through quasicrystalline confinement},\ }\href {https://doi.org/10.1103/PhysRevLett.133.196001} {\bibfield  {journal} {\bibinfo  {journal} {Phys. Rev. Lett.}\ }\textbf {\bibinfo {volume} {133}},\ \bibinfo {pages} {196001} (\bibinfo {year} {2024})}\BibitemShut {NoStop}%
\bibitem [{\citenamefont {Yu}\ \emph {et~al.}(2024)\citenamefont {Yu}, \citenamefont {Bhave}, \citenamefont {Reeve}, \citenamefont {Song},\ and\ \citenamefont {Schneider}}]{Yu2024}%
  \BibitemOpen
  \bibfield  {author} {\bibinfo {author} {\bibfnamefont {J.-C.}\ \bibnamefont {Yu}}, \bibinfo {author} {\bibfnamefont {S.}~\bibnamefont {Bhave}}, \bibinfo {author} {\bibfnamefont {L.}~\bibnamefont {Reeve}}, \bibinfo {author} {\bibfnamefont {B.}~\bibnamefont {Song}},\ and\ \bibinfo {author} {\bibfnamefont {U.}~\bibnamefont {Schneider}},\ }\bibfield  {title} {\bibinfo {title} {Observing the two-dimensional bose glass in an optical quasicrystal},\ }\href {https://doi.org/10.1038/s41586-024-07875-2} {\bibfield  {journal} {\bibinfo  {journal} {Nature}\ }\textbf {\bibinfo {volume} {633}},\ \bibinfo {pages} {338–343} (\bibinfo {year} {2024})}\BibitemShut {NoStop}%
\bibitem [{\citenamefont {Kempkes}\ \emph {et~al.}(2018)\citenamefont {Kempkes}, \citenamefont {Slot}, \citenamefont {Freeney}, \citenamefont {Zevenhuizen}, \citenamefont {Vanmaekelbergh}, \citenamefont {Swart},\ and\ \citenamefont {Smith}}]{Kempkes2018}%
  \BibitemOpen
  \bibfield  {author} {\bibinfo {author} {\bibfnamefont {S.~N.}\ \bibnamefont {Kempkes}}, \bibinfo {author} {\bibfnamefont {M.~R.}\ \bibnamefont {Slot}}, \bibinfo {author} {\bibfnamefont {S.~E.}\ \bibnamefont {Freeney}}, \bibinfo {author} {\bibfnamefont {S.~J.~M.}\ \bibnamefont {Zevenhuizen}}, \bibinfo {author} {\bibfnamefont {D.}~\bibnamefont {Vanmaekelbergh}}, \bibinfo {author} {\bibfnamefont {I.}~\bibnamefont {Swart}},\ and\ \bibinfo {author} {\bibfnamefont {C.~M.}\ \bibnamefont {Smith}},\ }\bibfield  {title} {\bibinfo {title} {Design and characterization of electrons in a fractal geometry},\ }\href {https://doi.org/10.1038/s41567-018-0328-0} {\bibfield  {journal} {\bibinfo  {journal} {Nature Physics}\ }\textbf {\bibinfo {volume} {15}},\ \bibinfo {pages} {127–131} (\bibinfo {year} {2018})}\BibitemShut {NoStop}%
\bibitem [{\citenamefont {Verstraten}\ \emph {et~al.}(2025)\citenamefont {Verstraten}, \citenamefont {Knottnerus}, \citenamefont {Tseng}, \citenamefont {Urech}, \citenamefont {Santo}, \citenamefont {Zampronio}, \citenamefont {Schreck}, \citenamefont {Spreeuw},\ and\ \citenamefont {Smith}}]{Verstraten2025}%
  \BibitemOpen
  \bibfield  {author} {\bibinfo {author} {\bibfnamefont {R.~C.}\ \bibnamefont {Verstraten}}, \bibinfo {author} {\bibfnamefont {I.~H.~A.}\ \bibnamefont {Knottnerus}}, \bibinfo {author} {\bibfnamefont {Y.~C.}\ \bibnamefont {Tseng}}, \bibinfo {author} {\bibfnamefont {A.}~\bibnamefont {Urech}}, \bibinfo {author} {\bibfnamefont {T.~S. d.~E.}\ \bibnamefont {Santo}}, \bibinfo {author} {\bibfnamefont {V.}~\bibnamefont {Zampronio}}, \bibinfo {author} {\bibfnamefont {F.}~\bibnamefont {Schreck}}, \bibinfo {author} {\bibfnamefont {R.~J.~C.}\ \bibnamefont {Spreeuw}},\ and\ \bibinfo {author} {\bibfnamefont {C.~M.}\ \bibnamefont {Smith}},\ }\href {https://doi.org/10.48550/ARXIV.2509.03514} {\bibinfo {title} {Control of single spin-flips in a rydberg atomic fractal}} (\bibinfo {year} {2025})\BibitemShut {NoStop}%
\end{thebibliography}%

\end{document}